# Exact QED Path Integration of the Maxwell Action, with Gravitational Curvature and Boundary Terms, Using Pontryagin Duality

*Jay R. Yablon*

**Abstract**

**We demonstrate how to explicitly calculate the QED path integral and associated Green functions, exactly, in curved spacetime, with retention of the boundary terms, to infinite order, for any and all spacetime manifolds with sufficient symmetry to admit the application of Pontryagin duality as a form of harmonic analysis. In the process we show how gauge symmetry itself greatly facilitates the ability to conduct harmonic analysis in curved spacetime and to do exact calculations with Pontryagin duality. We also show how non-Abelian, Yang-Mills gauge theories emerge naturally, if somewhat surprisingly, from this analysis.**

**Introduction**

In this paper, it is shown how to explicitly calculate the QED path integral, in curved spacetime, with retention of the boundary terms, exactly, and to infinite orders, for any and all spacetime manifolds with sufficient symmetry to admit the application of Pontryagin duality as a form of harmonic analysis. In sections 1 through 3, starting from the pure Maxwell Lagrangian density $\mathcal{L} = -\frac{1}{4} F_{\mu\nu} F^{\mu\nu}$ of classical electrodynamics, we integrate by parts the action $S(A^\mu) = \int_U \sqrt{-g}\, d^4 x\, \mathcal{L}$ in curved spacetime, including retention of the boundary term. After adding a source term $J^\mu A_\mu$ and a rest mass term $m^2$ in the usual way, we then quantize this action by calculating the QED path integral $Z = \int DA\, e^{iS}$, and we calculate the associated Green functions to infinite order.

What remains at that point is to obtain an explicit expression for the propagator in curved spacetime, which is the object of sections 4 through 7. In flat spacetime, this ordinarily requires Fourier analysis. But in curved spacetime, Fourier analysis no longer applies as is, and one must seek and apply suitable extensions and analogies which go under the broad title of "harmonic analysis." The ability to employ any particular form of harmonic analysis, is very dependent on the symmetries (or lack thereof) of the underlying spacetime manifold. Here, we explore harmonic analysis in curved spacetime using the particular analysis technique of *Pontryagin duality*, while restricting the spacetime manifolds which we consider to those for which this technique is applicable. We show, however, in particular, how gauge symmetry itself greatly facilitates the ability to conduct harmonic analysis in curved spacetime, and does allow exact calculations to be done with Pontryagin duality. We also show how non-Abelian, Yang-Mills gauge theories emerge naturally and somewhat surprisingly from this analysis.

**1. Integration by Parts of the Pure Maxwell Action in Curved Spacetime**

In a Riemannian manifold with gravitational interactions understood to originate from curved spacetime geometry, and applying the "minimal coupling" principle, the field strength tensor of classical electrodynamics is defined in relation to the vector gauge potential $A_\nu$ according to $F_{\mu\nu} \equiv \partial_{;\mu} A_\nu - \partial_{;\nu} A_\mu \equiv \partial_{;[\mu} A_{\nu]}$, where the covariant derivative is given by the well-known relationship $\partial_{;\mu} A_\nu = \partial_\mu A_\nu - \Gamma^\tau_{\mu\nu} A_\tau$ of differential geometry. Because the Christoffel connection $\Gamma^\tau_{\mu\nu} = \Gamma^\tau_{\nu\mu}$ is symmetric under transposition of its lower indexes, however, it is well known that the connection terms identically cancel in the particular antisymmetric combination used to form $F_{\mu\nu}$, so that:

$$F_{\mu\nu} = \partial_{;\mu} A_\nu - \partial_{;\nu} A_\mu = \partial_{;[\mu} A_{\nu]} = \partial_\mu A_\nu - \partial_\nu A_\mu = \partial_{[\mu} A_{\nu]}. \tag{1.1}$$

That is, in this particular antisymmetric combination of terms, the covariant derivatives may be replaced by ordinary partial derivatives (four-gradients).

To integrate the pure Maxwell action by parts, we start with the classical Lagrangian density for a pure, free electromagnetic field, and use (1.1) to write:

$$\mathcal{L} = -\tfrac{1}{4} F_{\mu\nu} F^{\mu\nu} = -\tfrac{1}{4} \partial_{;[\mu} A_{\nu]} \partial^{;[\mu} A^{\nu]} = -\tfrac{1}{4} \partial_{[\mu} A_{\nu]} \partial^{[\mu} A^{\nu]}. \tag{1.2}$$

We then use the flat spacetime product rule $\partial_\mu (A_\nu \partial^{[\mu} A^{\nu]}) = \partial_\mu A_\nu \partial^{[\mu} A^{\nu]} + A_\nu \partial_\mu \partial^{[\mu} A^{\nu]}$, from which we may easily form, by antisymmetric construction:

$$\mathcal{L} = -\tfrac{1}{4} F_{\mu\nu} F^{\mu\nu} = -\tfrac{1}{4} \partial_{[\mu} A_{\nu]} \partial^{[\mu} A^{\nu]} = -\tfrac{1}{4} \partial_{[\mu} (A_{\nu]} \partial^{[\mu} A^{\nu]}) + \tfrac{1}{4} A_{[\nu} \partial_{\mu]} \partial^{[\mu} A^{\nu]}. \tag{1.3}$$

We then make use of the minimal coupling principle to turn all of the ordinary derivatives back into covariant derivatives, so that the Lagrangian density in curved spacetime, or when using arbitrary coordinate systems, is:

$$\mathcal{L} = -\tfrac{1}{4}\partial_{:[\mu}\left(A_{\nu]}\partial^{:[\mu}A^{\nu]}\right) + \tfrac{1}{4}A_{[\nu}\partial_{:\mu]}\partial^{:[\mu}A^{\nu]} = -\tfrac{1}{2}\partial_{:\mu}\left(A_{\nu}\partial^{:[\mu}A^{\nu]}\right) + \tfrac{1}{2}A_{\mu}\left(g^{\mu\nu}\partial_{:\sigma}\partial^{:\sigma} - \partial^{:\nu}\partial^{:\mu}\right)A_{\nu}. \quad (1.4)$$

Here, it is apparent by inspection that the covariant derivatives do matter, and cannot be summarily treated as ordinary partial derivatives, if we wish to properly describe electrodynamics with gravitational curvature or in general coordinate systems. Most clearly, noting the term $\partial^{:\nu}\partial^{:\mu}A_{\nu}$ in the above, we recall that in differential geometry, the Riemann tensor $R^{\alpha}{}_{\beta\mu\nu}A_{\alpha} \equiv [\partial_{:\mu},\partial_{:\nu}]A_{\beta}$ is *defined* as a measure of the degree to which the *covariant* derivatives $\partial_{:\mu},\partial_{:\nu}$ may be commuted, based on an underlying analysis of the parallel transport of vectors, see, e.g., [1], section 6.5. Only where $R^{\alpha}{}_{\beta\mu\nu} = 0$, may we commute the derivatives using $[\partial_{:\mu},\partial_{:\nu}]A_{\beta} = [\partial_{\mu},\partial_{\nu}]A_{\beta} = 0$.

Thus, keeping in mind the term $\partial^{:\nu}\partial^{:\mu}A_{\nu}$ in (1.4), we write $R^{\alpha}{}_{\beta\mu\nu}A_{\alpha} = [\partial_{:\mu},\partial_{:\nu}]A_{\beta}$ as $R^{\beta\alpha}{}_{\mu\nu}A_{\alpha} = [\partial_{:\nu},\partial_{:\mu}]A^{\beta}$, using $R^{\alpha\beta}{}_{\mu\nu} = -R^{\beta\alpha}{}_{\mu\nu}$. Then, using $R^{\alpha}{}_{\mu} \equiv R^{\nu\alpha}{}_{\mu\nu}$, we contract $\nu,\beta$ to write $R^{\nu}{}_{\mu}A_{\nu} = [\partial_{:\nu},\partial_{:\mu}]A^{\nu}$, where $R^{\alpha}{}_{\mu}$ is the mixed Ricci (contracted Riemann) tensor. Finally, mindful that $R^{\mu\nu} = R^{\nu\mu}$ is symmetric under index transposition, we again juggle the indexes to finally write:

$$R^{\mu\nu}A_{\nu} = [\partial^{:\nu},\partial^{:\mu}]A_{\nu} = R^{\nu\mu}A_{\nu} = -[\partial^{:\mu},\partial^{:\nu}]A_{\nu}. \quad (1.5)$$

It is important that one of the derivatives $\partial^{:\nu}$ operates on and is contracted with a vector $A_{\nu}$ (or later, with an index of a higher rank tensor). Were one to remove $A_{\nu}$ from each side, one would get $R^{\mu\nu} = [\partial^{:\nu},\partial^{:\mu}]$, which in this stripped-down form would be invalid because a symmetric $R^{\mu\nu}$ would be set equal to an antisymmetric $[\partial^{:\nu},\partial^{:\mu}]$. The contracted equation (1.5), however, says that the *free index* in $R^{\mu\nu}$ – whether in the left or the right position in the symmetric $R^{\mu\nu}$ – is always on the right-hand derivative in the positively-specified commutator and the left-hand derivative in the negatively-specified commutator.

The above means that $\partial^{:\nu}\partial^{:\mu}A_{\nu} = \partial^{:\mu}\partial^{:\nu}A_{\nu} + R^{\mu\nu}A_{\nu}$, so we may take advantage of this to write (1.4) in the three alternative forms:

$$\begin{aligned}\mathcal{L} &= -\tfrac{1}{2}\partial_{:\mu}\left(A_{\nu}\partial^{:[\mu}A^{\nu]}\right) + \tfrac{1}{2}A_{\mu}\left(g^{\mu\nu}\partial_{:\sigma}\partial^{:\sigma} - \partial^{:\nu}\partial^{:\mu}\right)A_{\nu} \\ &= -\tfrac{1}{2}\partial_{:\mu}\left(A_{\nu}\partial^{:[\mu}A^{\nu]}\right) + \tfrac{1}{2}A_{\mu}\left(g^{\mu\nu}\partial_{:\sigma}\partial^{:\sigma} - \partial^{:\mu}\partial^{:\nu} + R^{\mu\nu}\right)A_{\nu} \\ &= -\tfrac{1}{2}\partial_{:\mu}\left(A_{\nu}\partial^{:[\mu}A^{\nu]}\right) + \tfrac{1}{2}A_{\mu}\left(g^{\mu\nu}\left(\partial_{:\sigma}\partial^{:\sigma} + \tfrac{1}{2}\kappa T\right) - \partial^{:\mu}\partial^{:\nu} - \kappa T^{\mu\nu}\right)A_{\nu}\end{aligned} \quad (1.6)$$

In the second line above, the Ricci curvature tensor $R^{\mu\nu}$ becomes an explicit term in the Lagrangian density. Then, in the third line, using the inverted form $R^{\mu\nu} = -\kappa(T^{\mu\nu} - \frac{1}{2}g^{\mu\nu}T)$ of the Einstein equation, the energy-momentum tensor $T^{\mu\nu}$ and the trace energy density $T$ appear, with the trace energy density $T$ moving naturally into the $g^{\mu\nu}(\partial_{;\sigma}\partial^{;\sigma} + \frac{1}{2}\kappa T)$ term of the Lagrangian density precisely where the rest mass is often situated, and with exactly the same sign. These terms will become central to quantum mechanical path integration of the electrodynamic action in a gravitational field.

Now, if $U$ represents the entire four-dimensional Riemannian manifold, then in flat spacetime, using the usual shorthand $d^4x \equiv dx^0 dx^1 dx^2 dx^3$, the action $S(A^\mu) = \int_U d^4x \mathcal{L}$. However, in curved spacetime, we must use the natural volume element $\sqrt{-g}\,d^4x$, which by Jacobi is required to maintain invariance under general coordinate transformations, see the original development of this at eq. (18) of [2], where the covariant (lower-indexed) metric tensor determinant $g = \frac{1}{4!}\varepsilon^{\mu\alpha\sigma\delta}\varepsilon^{\nu\beta\tau\gamma}g_{\mu\nu}g_{\alpha\beta}g_{\sigma\tau}g_{\delta\gamma}$, and $\varepsilon^{\mu\alpha\sigma\delta}$ is the totally-antisymmetric Levi-Civita tensor for which $\varepsilon^{0123} = 1$. Therefore, making use of (1.6) with an action $S(A^\mu) = \int_U \sqrt{-g}\,d^4x\,\mathcal{L}$, we write out three equivalent expressions for the electrodynamic action in curved spacetime:

$$\begin{aligned}
S(A^\mu) &= -\tfrac{1}{2}\int_U \sqrt{-g}\,d^4x \left[\partial_{;\mu}\left(A_\nu \partial^{:[\mu}A^{\nu]}\right)\right] + \int_U \sqrt{-g}\,d^4x\left[\tfrac{1}{2}A_\mu\left(g^{\mu\nu}\partial_{;\sigma}\partial^{;\sigma} - \partial^{;\nu}\partial^{;\mu}\right)A_\nu\right] \\
&= -\tfrac{1}{2}\int_U \sqrt{-g}\,d^4x\left[\partial_{;\mu}\left(A_\nu \partial^{:[\mu}A^{\nu]}\right)\right] + \int_U \sqrt{-g}\,d^4x\left[\tfrac{1}{2}A_\mu\left(g^{\mu\nu}\partial_{;\sigma}\partial^{;\sigma} - \partial^{;\mu}\partial^{;\nu} + R^{\mu\nu}\right)A_\nu\right] \quad (1.7) \\
&= -\tfrac{1}{2}\int_U \sqrt{-g}\,d^4x\left[\partial_{;\mu}\left(A_\nu \partial^{:[\mu}A^{\nu]}\right)\right] + \int_U \sqrt{-g}\,d^4x\left[\tfrac{1}{2}A_\mu\left(g^{\mu\nu}\left(\partial_{;\sigma}\partial^{;\sigma} + \tfrac{1}{2}\kappa T\right) - \partial^{;\mu}\partial^{;\nu} - \kappa T^{\mu\nu}\right)A_\nu\right]
\end{aligned}$$

The boundary term $-\tfrac{1}{2}\int_U \sqrt{-g}\,d^4x\left[\partial_{;\mu}\left(A_\nu \partial^{:[\mu}A^{\nu]}\right)\right]$ contains the covariant derivative $\partial_{;\mu}$, as well as $\sqrt{-g}$. But, if we define the vector $V^\mu \equiv A_\nu \partial^{:[\mu}A^{\nu]}$ and use the differential geometry identity $\partial_{;\mu}V^\mu = (1/\sqrt{-g})\partial_\mu(\sqrt{-g}\,V^\mu)$, we may simplify this by writing:

$$\partial_{;\mu}\left(A_\nu \partial^{:[\mu}A^{\nu]}\right) = \partial_{;\mu}V^\mu = \frac{1}{\sqrt{-g}}\partial_\mu\left(\sqrt{-g}\,V^\mu\right) = \frac{1}{\sqrt{-g}}\partial_\mu\left(\sqrt{-g}\,A_\nu \partial^{:[\mu}A^{\nu]}\right). \quad (1.8)$$

We can use this to in turn re-write, and then integrate, the boundary term. Thus:

$$\int_U \sqrt{-g}\,d^4x\left[\partial_{;\mu}\left(A_\nu \partial^{:[\mu}A^{\nu]}\right)\right] = \int_U d^4x\,\partial_\mu\left(\sqrt{-g}\,A_\nu \partial^{:[\mu}A^{\nu]}\right) = \int_{\partial U} (d^3x)_\mu \sqrt{-g}\,A_\nu \partial^{:[\mu}A^{\nu]}. \quad (1.9)$$

In the final equality, to integrate the boundary term, we apply Stokes' / Gauss' theorem $\int_U dH = \int_{\partial U} H$ for a differential $p$-form $H$, where $\partial U$ is the boundary of a $p+1$ dimensional manifold $U$, and where the three-volume element $(d^3x)_\mu = \frac{1}{3!}\varepsilon_{\mu\alpha\beta\sigma}dx^\alpha \wedge dx^\beta \wedge dx^\sigma$, showing explicit wedge products. In (1.9) above, we see that (1.8) kills three birds with one stone by simultaneously turning the covariant derivative $\partial_{;\mu}$ into an ordinary partial derivative $\partial_\mu$ which can be integrated, and by cancelling the $\sqrt{-g}$ factors and $1/\sqrt{-g}$ with one another outside of the $\partial_\mu$, and in the final term, by effectively moving the original $\sqrt{-g}$ into the after-integration expression $\sqrt{-g}A_\nu \partial^{;[\mu}A^{\nu]}$.

If we finally give the gauge field a small mass $\partial^\sigma \partial_\sigma \to \partial^\sigma \partial_\sigma + m^2$ "by hand" (rather than by the more "natural" approach of spontaneous symmetry breaking, and noting that in the third line of (1.7), the term $\frac{1}{2}\kappa T$ has already entered in a more "natural" way), and if we add a source term $A_\mu J^\mu$, then making use of (1.9), we can now rewrite (1.7) as:

$$\boxed{\begin{aligned}S(A^\mu) &= -\tfrac{1}{2}\int_{\partial U}\sqrt{-g}(d^3x)_\mu A_\nu \partial^{;[\mu}A^{\nu]} + \int_U \sqrt{-g}\,d^4x\left[\tfrac{1}{2}A_\mu\left(g^{\mu\nu}(\partial_{;\sigma}\partial^{;\sigma} + m^2) - \partial^{;\nu}\partial^{;\mu}\right)A_\nu + J^\nu A_\nu\right] \\ &= -\tfrac{1}{2}\int_{\partial U}\sqrt{-g}(d^3x)_\mu A_\nu \partial^{;[\mu}A^{\nu]} + \int_U \sqrt{-g}\,d^4x\left[\tfrac{1}{2}A_\mu\left(g^{\mu\nu}(\partial_{;\sigma}\partial^{;\sigma} + m^2) - \partial^{;\mu}\partial^{;\nu} + R^{\mu\nu}\right)A_\nu + J^\nu A_\nu\right] \\ &= -\tfrac{1}{2}\int_{\partial U}\sqrt{-g}(d^3x)_\mu A_\nu \partial^{;[\mu}A^{\nu]} + \int_U \sqrt{-g}\,d^4x\left[\tfrac{1}{2}A_\mu\left(g^{\mu\nu}(\partial_{;\sigma}\partial^{;\sigma} + \tfrac{1}{2}\kappa T + m^2) - \partial^{;\mu}\partial^{;\nu} - \kappa T^{\mu\nu}\right)A_\nu + J^\nu A_\nu\right]\end{aligned}} \quad (1.10)$$

This is the classical electrodynamic action, but in the presence of a gravitational field.

In the flat spacetime limit, where $g_{\mu\nu} \to \eta_{\mu\nu}$, and with $\text{diag}(\eta_{\mu\nu}) = (1,-1,-1,-1)$ in rectilinear coordinates, the connections $\Gamma^\alpha{}_{\mu\nu} = g^{\alpha\tau}(g_{\tau\mu,\nu} + g_{\nu\tau,\mu} - g_{\mu\nu,\tau}) \to 0$, the covariant derivatives become ordinary partial derivatives $\partial_{;\sigma} \to \partial_\sigma$ and thus commute as $[\partial_{;\mu}, \partial_{;\nu}] \to [\partial_\mu, \partial_\nu] = 0$, the Riemann and the Ricci tensors $R^\alpha{}_{\beta\mu\nu}A_\alpha = [\partial_{;\mu}, \partial_{;\nu}]A_\beta \to 0$ and $R^{\mu\nu}A_\nu = [\partial^{;\nu}, \partial^{;\mu}]A_\nu \to 0$, and the metric tensor determinant term $\sqrt{-g} \to 1$. Therefore, (1.10) reduces, in the flat spacetime limit, in rectilinear coordinates, to:

$$\begin{aligned}S(A^\mu) &= \int_{\partial U}(d^3x)_\mu A_\nu \partial^{[\mu}A^{\nu]} + \int_U d^4x\left[\tfrac{1}{2}A_\mu\left(g^{\mu\nu}(\partial_\sigma\partial^\sigma + m^2) - \partial^\nu\partial^\mu\right)A_\nu + J^\nu A_\nu\right] \\ &= \int_{\partial U}(d^3x)_\mu A_\nu \partial^{[\mu}A^{\nu]} + \int_U d^4x\left[\tfrac{1}{2}A_\mu\left(g^{\mu\nu}(\partial_\sigma\partial^\sigma + m^2) - \partial^\mu\partial^\nu\right)A_\nu + J^\nu A_\nu\right]\end{aligned} \quad (1.11)$$

If one takes $A_\nu$ to be zero over the boundary $\partial U$ as is often done, then the boundary term drops out, and (1.11) reduces simply to:

$$S(A^\mu) = \int_U d^4x \left[\tfrac{1}{2} A_\mu \left(g^{\mu\nu}(\partial_\sigma \partial^\sigma + m^2) - \partial^\nu \partial^\mu\right) A_\nu + J^\nu A_\nu \right] = \int_U d^4x \left[\tfrac{1}{2} A_\mu \left(g^{\mu\nu}(\partial_\sigma \partial^\sigma + m^2) - \partial^\mu \partial^\nu\right) A_\nu + J^\nu A_\nu \right]. \quad (1.12)$$

Equation (1.12), with the boundary term discarded, is often used in a well-known way to arrive at the familiar momentum space propagator $D_{\alpha\nu}(k) = \left(-g_{\alpha\nu} + k_\alpha k_\nu / m^2\right)/\left(k_\sigma k^\sigma - m^2\right)$ of QED. (A very good treatment of this is given in Chapter I.5 of [3].) Using this known calculation as a template, we shall explore how to extend this approach to path integrate (1.10) in curved spacetime, without discarding the boundary term.

## 2. QED Path Integration in Curved Spacetime with the Boundary Term Retained

Let us now start to calculate the path integral $Z = \int DA e^{iS}$ using the action (1.10). At the outset, to work with an "apples to apples" expression, we place all terms under the same integral over the four-volume element $\sqrt{-g} d^4x$, using Gauss' / Stokes' theorem in (1.9) to take a step back. Thus:

$$\begin{aligned} Z = \int DA e^{iS(A^\mu)} &\equiv \mathcal{C} \exp i[W(J)] \equiv \\ &= \int DA \exp i \int_U d^4x \left[-\tfrac{1}{2} \partial_\mu \left(\sqrt{-g} A_\nu \partial^{[\mu} A^{\nu]}\right) + \sqrt{-g} \tfrac{1}{2} A_\mu \left(g^{\mu\nu}(\partial_{;\sigma} \partial^{;\sigma} + m^2) - \partial^{;\nu}\partial^{;\mu}\right) A_\nu + \sqrt{-g} J^\nu A_\nu \right] \\ &= \int DA \exp i \int_U d^4x \left[-\tfrac{1}{2} \partial_\mu \left(\sqrt{-g} A_\nu \partial^{[\mu} A^{\nu]}\right) + \sqrt{-g} \tfrac{1}{2} A_\mu \left(g^{\mu\nu}(\partial_{;\sigma} \partial^{;\sigma} + m^2) - \partial^{;\mu}\partial^{;\nu} + R^{\mu\nu}\right) A_\nu + \sqrt{-g} J^\nu A_\nu \right] \\ &= \int DA \exp i \int_U d^4x \left[-\tfrac{1}{2} \partial_\mu \left(\sqrt{-g} A_\nu \partial^{[\mu} A^{\nu]}\right) + \sqrt{-g} \tfrac{1}{2} A_\mu \left(g^{\mu\nu}(\partial_{;\sigma} \partial^{;\sigma} + \tfrac{1}{2}\kappa T + m^2) - \partial^{;\mu}\partial^{;\nu} - \kappa T^{\mu\nu}\right) A_\nu + \sqrt{-g} J^\nu A_\nu \right] \end{aligned} \quad .(2.1)$$

Now, all we need to do is calculate $W(J)$ defined above, which, of course, is not a trivial task, especially in curved spacetime, and which will consume the balance of this paper.

Solutions to this path integral all emanate from the basic Gaussian mathematical identity:

$$\int dx \exp i \left[\tfrac{1}{2} ax^2 + Jx\right] = \left(\frac{2\pi i}{a}\right)^{.5} \exp i \left[-\frac{1}{2} \frac{J^2}{a}\right], \quad (2.2)$$

in progressively complicated variations. The first step up is to rescale $a \to \sqrt{g_{00}} a$ and $J \to \sqrt{g_{00}} J$, to lay an eventual foundation for the term $\sqrt{-g}$ which arises when the spacetime manifold is curved and / or one makes an arbitrary coordinate choice. Thus, we write:

$$\int dx \exp i \left[\tfrac{1}{2} \sqrt{g_{00}} ax^2 + \sqrt{g_{00}} Jx\right] = \left(\frac{2\pi i}{\sqrt{g_{00}} a}\right)^{.5} \exp i \left[-\sqrt{g_{00}} \frac{1}{2} \frac{J^2}{a}\right] \equiv \mathcal{C} \exp i \left[-\sqrt{g_{00}} \frac{1}{2} \frac{J^2}{a}\right]. \quad (2.3)$$

From here on, we will not need to be concerned with the overall factor $\mathcal{C}$ which emanates from infinite products $\prod_{n=1}^{\infty}$ of variants of the term $(2\pi i/a)^5$, and shall remain focused on the exponential terms. Next, we move up to:

$$\int D\varphi \exp i\left[\tfrac{1}{2}\sqrt{-g}\,\varphi \cdot K \cdot \varphi + \sqrt{-g}\,J \cdot \varphi\right] = \mathcal{C}\exp i\left[-\tfrac{1}{2}\sqrt{-g}\,J \cdot K^{-1} \cdot J\right]. \tag{2.4}$$

Now, keeping in mind the earlier scaling $J \to \sqrt{g_{00}}\,J$, let us operate on the exponential on the left hand side of (2.4) with the functional derivative $-i\delta/\delta(\sqrt{-g}\,J)$, to obtain:

$$-i\frac{\delta}{\delta(\sqrt{-g}\,J)}\exp i\left[\tfrac{1}{2}\sqrt{-g}\,\varphi \cdot K \cdot \varphi + \sqrt{-g}\,J \cdot \varphi\right] = \varphi \exp i\left[\tfrac{1}{2}\sqrt{-g}\,\varphi \cdot K \cdot \varphi + \sqrt{-g}\,J \cdot \varphi\right], \tag{2.5}$$

which enables, in context, the substitution $\varphi \to -i\delta/\delta(\sqrt{-g}\,J)$. Inserting a term $-V(\phi)$ into the exponent on the left hand side of (2.4) and applying (2.5) then allows us to take a critical step forward by writing (see, e.g., section I.7 of [3]):

$$\int D\varphi \exp i\left[-V(\phi) + \tfrac{1}{2}\sqrt{-g}\,\varphi \cdot K \cdot \varphi + \sqrt{-g}\,J \cdot \varphi\right]$$
$$= \int D\varphi \exp i(-V(\phi))\exp i\left[\tfrac{1}{2}\sqrt{-g}\,\varphi \cdot K \cdot \varphi + \sqrt{-g}\,J \cdot \varphi\right]$$
$$= \int D\varphi \exp i\left(-V\left(-i\frac{\delta}{\delta(\sqrt{-g}\,J)}\right)\right)\exp i\left[+\tfrac{1}{2}\sqrt{-g}\,\varphi \cdot K \cdot \varphi + \sqrt{-g}\,J \cdot \varphi\right]. \tag{2.6}$$
$$= \exp i\left(-V\left(-i\frac{\delta}{\delta(\sqrt{-g}\,J)}\right)\right)\int D\varphi \exp i\left[+\tfrac{1}{2}\sqrt{-g}\,\varphi \cdot K \cdot \varphi + \sqrt{-g}\,J \cdot \varphi\right]$$
$$= \mathcal{C}\exp i\left(-V\left(-i\frac{\delta}{\delta(\sqrt{-g}\,J)}\right)\right)\exp i\left[-\tfrac{1}{2}\sqrt{-g}\,J \cdot K^{-1} \cdot J\right]$$

In the second line above, we merely split the exponential. In the third line, we employ the substitution $\varphi \to -i\delta/\delta(\sqrt{-g}\,J)$ developed in (2.5). In the fourth line, we can remove the term $=\exp i(-V(-i\delta/\delta(\sqrt{-g}\,J)))$ to the outside of the path integral over $D\varphi$, because this term is no longer a function of the integration variable $\varphi$. In the final line, we simply employ (2.4).

Finally, we step (2.6) up to the integral variant, taken locally:

$$\int D\varphi \exp i\int d^4x\left[-V(\phi) + \tfrac{1}{2}\sqrt{-g(x)}\,\varphi(x) \cdot K \cdot \varphi(x) + \sqrt{-g(x)}\,J(x) \cdot \varphi(x)\right]$$
$$= \mathcal{C}\exp i\int d^4x\left(-V\left(-i\frac{\delta}{\delta(\sqrt{-g}\,J)}\right)\right)\exp i\int d^4x\left[-\tfrac{1}{2}\sqrt{-g(x)}\,J(x) \cdot K^{-1} \cdot J(x)\right] = \mathcal{C}\exp iW(J(x)) \tag{2.7}$$

The top line above clearly mirrors (2.1). Contrasting (2.1) term-by-term with (2.7), we can pick off the associations (" $\leftrightarrow$ "):

$$\varphi \leftrightarrow A, \tag{2.8}$$

$$V(\phi) \leftrightarrow \tfrac{1}{2}\partial_\mu\left(\sqrt{-g}A_\nu \partial^{:[\mu}A^{\nu]}\right) \leftrightarrow V\left(-i\delta/\delta(\sqrt{-g}J)\right), \tag{2.9}$$

$$\begin{aligned} K &\leftrightarrow g^{\mu\nu}\left(\partial_{;\sigma}\partial^{;\sigma}+m^2\right)-\partial^{:\nu}\partial^{:\mu} \\ &\leftrightarrow g^{\mu\nu}\left(\partial_{;\sigma}\partial^{;\sigma}+m^2\right)-\partial^{:\mu}\partial^{:\nu}+R^{\mu\nu} \\ &\leftrightarrow g^{\mu\nu}\left(\partial_{;\sigma}\partial^{;\sigma}+\tfrac{1}{2}\kappa T+m^2\right)-\partial^{:\mu}\partial^{:\nu}-\kappa T^{\mu\nu} \end{aligned}, \tag{2.10}$$

$$\begin{aligned} K^{-1} &\leftrightarrow \left(g^{\mu\nu}\left(\partial_{;\sigma}\partial^{;\sigma}+m^2\right)-\partial^{:\nu}\partial^{:\mu}\right)^{-1} \\ &\leftrightarrow \left(g^{\mu\nu}\left(\partial_{;\sigma}\partial^{;\sigma}+m^2\right)-\partial^{:\mu}\partial^{:\nu}+R^{\mu\nu}\right)^{-1} \\ &\leftrightarrow \left(g^{\mu\nu}\left(\partial_{;\sigma}\partial^{;\sigma}+\tfrac{1}{2}\kappa T+m^2\right)-\partial^{:\mu}\partial^{:\nu}-\kappa T^{\mu\nu}\right)^{-1} \end{aligned}. \tag{2.11}$$

Now, let's develop the boundary / $V(\phi)$ term in (2.9) a bit more. Here, we start with the substitution $\varphi \to -i\delta/\delta(\sqrt{-g}J)$ developed in (2.5) to substitute $A^{\nu]} \to -i\delta/\delta(\sqrt{-g}J_{\nu]})$ and $A_\nu \to -i\delta/\delta(\sqrt{-g}J^\nu)$ in (2.9). Thus, defining and using the shorthand $\delta_\nu \equiv \delta/\delta(\sqrt{-g}J^\nu)$, we replace (2.9) with the more specific:

$$V(\phi) \leftrightarrow \tfrac{1}{2}\partial_\mu\left(\sqrt{-g}A_\nu \partial^{:[\mu}A^{\nu]}\right) \leftrightarrow -\tfrac{1}{2}\partial_\mu\left(\sqrt{-g}\frac{\delta}{\delta(\sqrt{-g}J^\nu)}\partial^{:[\mu}\frac{\delta}{\delta(\sqrt{-g}J_{\nu]})}\right) \equiv -\tfrac{1}{2}\partial_\mu\left(\sqrt{-g}\delta_\nu \partial^{:[\mu}\delta^{\nu]}\right). \tag{2.12}$$

Putting (2.7) and (2.1) together using (2.8) through (2.12), finally allows us to write down an exact solution to the path integral in (2.1), in terms of the energy / momentum tensor and its trace, using the final of the three alternative expressions in (2.1), as follows:

$$\begin{aligned} &\int D\varphi \exp i\int d^4x\left[-\tfrac{1}{2}\partial_\mu\left(\sqrt{-g}A_\nu\partial^{:[\mu}A^{\nu]}\right)+\tfrac{1}{2}\sqrt{-g}A_\mu\left(g^{\mu\nu}\left(\partial_{;\sigma}\partial^{;\sigma}+\tfrac{1}{2}\kappa T+m^2\right)-\partial^{:\mu}\partial^{:\nu}-\kappa T^{\mu\nu}\right)A_\nu+\sqrt{-g}J^\nu A_\nu\right] \\ &= \mathcal{C}\exp i\int d^4x\left(\tfrac{1}{2}\partial_\mu\left(\sqrt{-g}\delta_\nu\partial^{:[\mu}\delta^{\nu]}\right)\right)\exp i\int d^4x\left[-\tfrac{1}{2}\sqrt{-g}J^\mu\left(g^{\mu\nu}\left(\partial_{;\sigma}\partial^{;\sigma}+\tfrac{1}{2}\kappa T+m^2\right)-\partial^{:\mu}\partial^{:\nu}-\kappa T^{\mu\nu}\right)^{-1}J^\nu\right] \\ &= \mathcal{C}\exp i\int \sqrt{-g}\left(d^3x\right)_\mu\left[\tfrac{1}{2}\delta_\nu\partial^{:[\mu}\delta^{\nu]}\right]\exp i\int \sqrt{-g}d^4x\left[-\tfrac{1}{2}J^\mu\left(g^{\mu\nu}\left(\partial_{;\sigma}\partial^{;\sigma}+\tfrac{1}{2}\kappa T+m^2\right)-\partial^{:\mu}\partial^{:\nu}-\kappa T^{\mu\nu}\right)^{-1}J^\nu\right] \\ &= \mathcal{C}\exp iW(J(x)) \end{aligned} \tag{2.13}$$

In the third line, we have used Gauss' / Stokes' theorem to integrate the boundary term. We note as an aside that the first line contains the term $A_\mu \partial^{:\mu}\partial^{:\nu}A_\nu$, and, given the need to exercise care with the non commuting $\partial^{:\mu}\partial^{:\nu}A_\nu$, that we have maintained the exact same index configuration in the $J^\mu\left(\partial^{:\mu}\partial^{:\nu}\right)^{-1}J^\nu$ portion of the second and third lines.

The above contains an "inverse" of $g^{\mu\nu}(\partial_{;\sigma}\partial^{;\sigma} + \frac{1}{2}\kappa T + m^2) - \partial^{;\mu}\partial^{;\nu} - \kappa T^{\mu\nu}$. Of course, we still need to find an explicit expression for this inverse. To begin this process, we treat this inverse *non-locally*, and so *define* this inverse as the spacetime propagator $D_{\nu\alpha}(x-y)$, using the various formulations which appear in (2.1), and using the "unit" Kronecker $\delta^\mu{}_\alpha$ times the "unit" Dirac four-delta $\delta^{(4)}(x)$, according to the $D_{\nu\alpha}(x-y)$ definition:

$$\begin{aligned}
\delta^\mu{}_\alpha \delta^{(4)}(x-y) &\equiv \left(g^{\mu\nu}(\partial_{;\sigma}\partial^{;\sigma} + m^2) - \partial^{;\nu}\partial^{;\mu}\right)D_{\nu\alpha}(x-y) \\
&= \left(g^{\mu\nu}(\partial_{;\sigma}\partial^{;\sigma} + m^2) - \partial^{;\mu}\partial^{;\nu} + R^{\mu\nu}\right)D_{\nu\alpha}(x-y) \\
&= \left(g^{\mu\nu}(\partial_{;\sigma}\partial^{;\sigma} + \tfrac{1}{2}\kappa T + m^2) - \partial^{;\mu}\partial^{;\nu} - \kappa T^{\mu\nu}\right)D_{\nu\alpha}(x-y)
\end{aligned} \qquad (2.14)$$

Recall now, from (1.5), that $R^{\mu\nu}A_\nu = [\partial^{;\nu}, \partial^{;\mu}]A_\nu \neq 0$, and that the inverses in (2.13) which are used to define $D_{\nu\alpha}(x-y)$ in (2.14) all contain the non-commuting term $\partial^{;\nu}\partial^{;\mu}$ or $\partial^{;\mu}\partial^{;\nu}$ together the transposition-symmetric $g^{\mu\nu}$, $R^{\mu\nu}$, and / or $T^{\mu\nu}$. Because of this – which originates in parallel transport analysis – *we cannot assume in curved spacetime that* $D_{\nu\alpha}(x-y) = D_{\alpha\nu}(x-y)$. In fact, *we are required to assume that in curved spacetime*, $D_{\nu\alpha}(x-y) \neq D_{\alpha\nu}(x-y)$ ! Thus, we will need to pay attention to the order of indexes in the propagator and not simply commute these at will. That is why we were careful just above in (2.13) with matching $A_\mu \partial^{;\mu}\partial^{;\nu} A_\nu$ to $J^\mu (\partial^{;\mu}\partial^{;\nu})^{-1} J^\nu$.

With all of the foregoing, we now return to (2.13), and write the amplitude function $W(J(x))$ directly as using $D_{\mu\nu}(x-y)$ as:

$$\boxed{\begin{aligned}
Z &= \mathcal{C} \exp iW(J(x)) \\
&= \mathcal{C} \exp i \int_{\partial U} \sqrt{-g}\,(d^3x)_\mu \left[\tfrac{1}{2}\delta_\nu \partial^{;[\mu}\delta^{\nu]}\right] \int_U \exp i \int \sqrt{-g}\,d^4x \sqrt{-g}\,d^4y \left[-\tfrac{1}{2} J^\mu(x) D_{\mu\nu}(x-y) J^\nu(y)\right]
\end{aligned}} \quad (2.15)$$

*The above fully solves the path integral (2.1) in curved spacetime, in principle, while retaining rather than discarding the boundary term.*

What remains is to calculate an explicit expression for $W(J)$. This in turn requires two things: First, in the next section, we shall develop the Wick contraction / Green functions related to the above $W(J(x))$. Thereafter, in order to obtain an explicit expression for $D_{\nu\alpha}(x-y)$, we will need to closely consider harmonic analysis and what happens to flat-spacetime Fourier analysis, in curved spacetime, including imposing certain symmetry restrictions on the spacetime manifold in order to permit such analysis to be done.

## 3. Wick Differentiation and Derivation of the Curved Spacetime QED Green Functions

Starting from the path integral (2.15), in order to work with an "apples-to-apples" expression in which all integrations are taken over a four-volume, we will find it helpful to again step back via Gauss' / Stokes' theorem $\int_{\partial U}(d^3x)_\mu \leftrightarrow \int_U d^4x \partial_\mu$ and return to a boundary term expressed as a four-volume integral. Thus we rewrite (2.15) as:

$$Z = \mathcal{C} \exp iW(J(x))$$
$$= \mathcal{C} \exp i \int_U \sqrt{-g}\, d^4x \partial_\mu \left[\tfrac{1}{2}\delta_\nu \partial^{:[\mu}\delta^{\nu]}\right] \int_U \exp i \int \sqrt{-g}\, d^4x \sqrt{-g}\, d^4y \left[-\tfrac{1}{2}J^\mu(x)D_{\mu\nu}(x-y)J^\nu(y)\right]. \quad (3.1)$$

As a preliminary matter, we deconstruct (3.1) into a local formulation without integrals:

$$Z = \mathcal{C} \exp iW(J) = \mathcal{C} \exp i\sqrt{-g}\, \partial_\mu \left[\tfrac{1}{2}\delta_\nu \partial^{:[\mu}\delta^{\nu]}\right] \exp i\left[-\tfrac{1}{2}\sqrt{-g}\, J^\mu D_{\mu\nu}\sqrt{-g}\, J^\nu\right]. \quad (3.2)$$

The lowest order-series expansion of the boundary term on the left is:

$$\exp i\sqrt{-g}\, \partial_\mu \left[\tfrac{1}{2}\delta_\nu \partial^{:[\mu}\delta^{\nu]}\right] \cong 1 + i\sqrt{-g}\, \partial_\mu \left[\tfrac{1}{2}\delta_\nu \partial^{:[\mu}\delta^{\nu]}\right] + \ldots, \quad (3.3)$$

while that for the right hand term this series is:

$$\exp i\left[-\tfrac{1}{2}\sqrt{-g}\, J^\alpha D_{\alpha\beta}\sqrt{-g}\, J^\beta\right] \cong 1 - \tfrac{1}{2}i\sqrt{-g}\, J^\alpha D_{\alpha\beta}\sqrt{-g}\, J^\beta + \ldots. \quad (3.4)$$

Now, we use the lowest order term $i\sqrt{-g}\, \partial_\mu \left[\tfrac{1}{2}\delta_\nu \partial^{:[\mu}\delta^{\nu]}\right]$ in (3.3) to operate on the entirety of $\exp i\left[-\tfrac{1}{2}\sqrt{-g}\, J^\mu D_{\mu\nu}\sqrt{-g}\, J^\nu\right]$ in (3.2). Mindful that $\delta^\nu \equiv \delta/\delta(\sqrt{-g}\, J_\nu)$, so that, for example, $\delta^{\nu]}(\sqrt{-g}\, J_\sigma) \equiv \delta^{\nu]}{}_\sigma$, we obtain the following for the first functional differentiation with $\delta^{\nu]}$:

$$i\sqrt{-g}\, \partial_\mu \left[\tfrac{1}{2}\delta_\nu \partial^{:[\mu}\delta^{\nu]}\right] \exp i\left[-\tfrac{1}{2}\sqrt{-g}\, J^\alpha D_{\alpha\beta}\sqrt{-g}\, J^\beta\right]$$
$$= \sqrt{-g}\, \partial_\mu \left[\tfrac{1}{2}\delta_\nu \partial^{:[\mu}\left[\tfrac{1}{2}D^{\{\nu|\beta\}}\sqrt{-g}\, J_\beta\right]\exp i\left[-\tfrac{1}{2}\sqrt{-g}\, J^\alpha D_{\alpha\beta}\sqrt{-g}\, J^\beta\right]. \quad (3.5)$$

We keep the factor of $\tfrac{1}{2}$ pinned to the anticommutator $D^{\{\nu\beta\}}$, because for the special case of a commuting $D^{\nu\beta} = D^{\beta\nu}$, we have $\tfrac{1}{2}D^{\{\nu\beta\}} = D^{\nu\beta}$.

The second functional differential $\delta_\nu$ is applied serially to the whole result in (3.5), that is, we must take $\delta\delta J = \delta^2 J$, not $\delta\delta J = (\delta J)^2$. Thus, proceeding from (3.5), we next obtain:

$$i\sqrt{-g}\partial_\mu \left[\tfrac{1}{2}\delta_\nu \partial^{:[\mu}\delta^{\nu]}\right] \exp i\left[-\tfrac{1}{2}\sqrt{-g}J^\alpha D_{\alpha\beta}\sqrt{-g}J^\beta\right]$$
$$=\sqrt{-g}\partial_\mu \tfrac{1}{2}\delta_\nu \left[\partial^{:[\mu}\tfrac{1}{2}D^{\{\nu\}\beta\}}\sqrt{-g}J_\beta \exp i\left(-\tfrac{1}{2}\sqrt{-g}J^\alpha D_{\alpha\beta}\sqrt{-g}J^\beta\right)\right] \qquad (3.6)$$
$$=\sqrt{-g}\partial_\mu \left[\tfrac{1}{2}\left(g_{\beta\nu}-i\tfrac{1}{2}D_{\{\nu\alpha\}}\sqrt{-g}J^\alpha \sqrt{-g}J_\beta\right)\partial^{:[\mu}\tfrac{1}{2}D^{\{\nu\}\beta\}}\right]\exp i\left[-\tfrac{1}{2}\sqrt{-g}J^\alpha D_{\alpha\beta}\sqrt{-g}J^\beta\right]$$

In the above, the term with $g_{\beta\nu}$ in the final line above, operating on the first order expansion term $-\tfrac{1}{2}i\sqrt{-g}J^\alpha D_{\alpha\beta}\sqrt{-g}J^\beta$ in (3.4), will be of interest later on. We write this "term of interest" as:

$$\sqrt{-g}\partial_\mu \left(\tfrac{1}{2}g_{\beta\nu}\partial^{:[\mu}\tfrac{1}{2}D^{\{\nu\}\beta\}}\right)\left(-\tfrac{1}{2}i\sqrt{-g}J^\alpha D_{\alpha\beta}\sqrt{-g}J^\beta\right)$$
$$=-i\sqrt{-g}\partial_\mu \left(\tfrac{1}{2}\partial^{:[\mu}\tfrac{1}{2}D^{\{\nu\}}{}_{\nu\}}\right)\left(\tfrac{1}{2}\sqrt{-g}J^\alpha D_{\alpha\beta}\sqrt{-g}J^\beta\right) \qquad (3.7)$$

With these preliminaries completed, we turn to calculating the Green functions. In order to calculate Green functions for (3.1) and it local cousin (3.2), we will further deconstruct (3.2) into it simplest structural form, and then reconstruct everything once the full calculation is complete. Thus, we deconstruct the amplitude in (3.2) into the following form:

$$\exp iW(J(x)) = \exp i\sqrt{-g}\partial_\mu \left[\tfrac{1}{2}\delta_\nu \partial^{:[\mu}\delta^{\nu]}\right]\exp i\left[-\tfrac{1}{2}\sqrt{-g}J^\mu D_{\mu\nu}\sqrt{-g}J^\nu\right]$$
$$\to \exp\left(\tfrac{1}{2}id\,\partial\,\delta^2\right)\exp\left(-\tfrac{1}{2}iDJ^2\right) \qquad (3.8)$$

In the above, we have made the symbolic deconstructions:

$$\sqrt{-g}\partial_\mu \to d, \qquad (3.9)$$
$$\delta_\nu \delta^{\nu]} \to \delta^2, \qquad (3.10)$$
$$\partial^{:[\mu} \to \partial, \qquad (3.11)$$
$$\sqrt{-g}J^\mu \sqrt{-g}J^\mu \to J^2, \qquad (3.12)$$
$$D_{\mu\nu} \to D. \qquad (3.13)$$

We use $\sqrt{-g}\partial_\mu \to d$ in (3.9) as a reminder that this is the derivative which gets converted via the general Gauss / Stokes conversion $\int_{\partial U}(d^3x)_\mu \leftrightarrow \int_U d^4x \partial_\mu$, and we hide the indexes, commutators, commutation position, and brackets with the understanding that these can later be reconstructed using (3.8) through (3.13) in the opposite direction to guide the proper replacement of the foregoing. Then, since (3.8) is a symbolic representation of (3.2), we can reconstruct the local (3.2) into the integral, non-local form (3.1). Finally, it will be an extremely useful aid in later reconstruction, to deconstruct (3.7) into the shorthand form:

$$-i\sqrt{-g}\partial_\mu\left(\tfrac{1}{2}\partial^{:[\mu}\tfrac{1}{2}D^{\{\nu]}{}_{\nu\}}\right)\left(\tfrac{1}{2}\sqrt{-g}J^\alpha D_{\alpha\beta}\sqrt{-g}J^\beta\right) \to -\tfrac{1}{8}i(d\partial D)(DJ^2). \tag{3.14}$$

Now, let us follow the Wick procedure by repeatedly differentiating using $\delta = \delta/\delta J$ in (3.8). Both of the exponentials in (3.8), $\exp\tfrac{1}{2}id\partial\delta^2$ and $\exp(-\tfrac{1}{2}iDJ^2)$, are of the general form $\exp\tfrac{1}{2}DJ^2$, and can be readily obtained from $\exp\tfrac{1}{2}DJ^2$ by re-scaling and / or renaming $D$. So, we repeatedly apply $\delta = \delta/\delta J$ to $\exp\tfrac{1}{2}DJ^2$, evaluate each successive derivative at $J=0$, and then reconstruct $\exp\tfrac{1}{2}DJ^2$ using a Maclaurin series. The first functional derivative, obtained and then evaluated at $J=0$ (denoted $\delta(0)$) is:

$$\delta\exp\tfrac{1}{2}DJ^2 = DJ\exp\tfrac{1}{2}DJ^2 \to \delta(0)=0. \tag{3.15}$$

The second functional derivative is:

$$\delta^2\exp\tfrac{1}{2}DJ^2 = \delta(DJ\exp\tfrac{1}{2}DJ^2) = [D+D^2J^2]\exp\tfrac{1}{2}DJ^2 \to \delta^2(0)=D. \tag{3.16}$$

Repeating this out to the eighth functional derivative, which will be sufficient to establish a pattern for an infinite summation, we find, successively:

$$\delta^3\exp\tfrac{1}{2}DJ^2 = [3D^2J+D^3J^3]\exp\tfrac{1}{2}DJ^2 \to \delta^3(0)=0, \tag{3.17}$$

$$\delta^4\exp\tfrac{1}{2}DJ^2 = [3D^2+6D^3J^2+D^4J^4]\exp\tfrac{1}{2}DJ^2 \to \delta^4(0)=3D^2, \tag{3.18}$$

$$\delta^5\exp\tfrac{1}{2}DJ^2 = [15D^3J+10D^4J^3+D^5J^5]\exp\tfrac{1}{2}DJ^2 \to \delta^5(0)=0, \tag{3.19}$$

$$\delta^6\exp\tfrac{1}{2}DJ^2 = [15D^3+45D^4J^2+15D^5J^4+D^6J^6]\exp\tfrac{1}{2}DJ^2 \to \delta^6(0)=15D^3, \tag{3.20}$$

$$\delta^7\exp\tfrac{1}{2}DJ^2 = [105D^4J+105D^5J^3+21D^6J^5+D^7J^7]\exp\tfrac{1}{2}DJ^2 \to \delta^7(0)=0, \tag{3.21}$$

$$\delta^8\exp\tfrac{1}{2}DJ^2 = [105D^4+420D^5J^2+210D^6J^4+28D^7J^6+D^8J^8]\exp\tfrac{1}{2}DJ^2 \to \delta^8(0)=105D^4. \tag{3.22}$$

We may summarize all of the above by:

$$\begin{cases}\delta^n(0)=(n-1)!!D^{\frac{n}{2}} & \text{for } n \text{ even}\\ \delta^n(0)=0 & \text{for } n \text{ odd}\end{cases}, \tag{3.23}$$

Using this to construct a Maclaurin series, we now obtain:

$$\begin{aligned}\exp\frac{1}{2}DJ^2 &= 1+\frac{1!!D}{2!}J^2+\frac{3!!D^2}{4!}J^4+\frac{5!!D^3}{6!}J^6+\frac{7!!D^4}{8!}J^8+\ldots\\ &= 1+\frac{1}{2!!}DJ^2+\frac{1}{4!!}D^2J^4+\frac{1}{6!!}D^3J^6+\frac{1}{8!!}D^4J^8+\ldots\\ &= \sum_{n=0}^\infty \frac{1}{2n!!}D^n J^{2n}\end{aligned} \tag{3.24}$$

In the foregoing, we come across the double-factorials which are endemic to Wick contractions and quantum field theory. Indeed, if we take note of the usual "hyperbolic cosine" function $\cosh x$, we see that the only difference between (3.24) and the ordinary cosh function is that the single factorial $2n!$ is replaced by the double factorial $2n!!$. This observation reveals a whole system of "double-factorial" mathematical functions, which we shall refer to as "Wick functions." For example, by analogy with and in contrast to the ordinary cosh function, we define, for example, a "coshw" function, the "Wick hyperbolic cosine," according to:

$$\begin{cases} \coshw(x) \equiv \sum_{n=0}^{\infty} \frac{1}{2n!!} x^{2n} = \exp\frac{1}{2}x^2 \\ \cosh(x) \equiv \sum_{n=0}^{\infty} \frac{1}{2n!} x^{2n} \end{cases}. \quad (3.25)$$

More generally, it is helpful think about all of this in the purely mathematical context of "Wick-Maclaurin functions," denoted by $fw(x)$, with the series expansion *defined*, and contrasted to the that of the usual Maclaurin series, such that:

$$fw(x) \equiv 1 + f'x + \frac{1}{2!!} f'' x^2 + \frac{1}{3!!} f''' x^3 + \frac{1}{4!!} f^{(4)} x^4 + \ldots$$
$$f(x) = 1 + f'x + \frac{1}{2!} f'' x^2 + \frac{1}{3!} f''' x^3 + \frac{1}{4!} f^{(4)} x^4 + \ldots \quad (3.26)$$

Thus, we see that every ordinary mathematical function has an analogous "Wick function" in which the single factorials in the series expansions are simply replaced with double factorials. Such functions are the same as their ordinary functional counterparts up to $x^2$ order (because $2! = 2!!$), but begin to have different characteristics beyond second order.

Now, we return to (3.8). Using (3.25) and (3.26), we may rescale and rename to express each of $\exp(\tfrac{1}{2}id\partial\delta^2)$ and $\exp(-\tfrac{1}{2}iDJ^2)$ in (3.8) as:

$$\exp(\tfrac{1}{2}id\partial\delta^2) = \coshw(\sqrt{i}\sqrt{d\partial}\delta) = \sum_{n=0}^{\infty} \frac{1}{2n!!} i^n (d\partial)^n \delta^{2n}, \quad (3.27)$$

$$\exp(-\tfrac{1}{2}iDJ^2) = \coshw(i\sqrt{i}\sqrt{D}J) = \sum_{n=0}^{\infty} \frac{1}{2n!!} (-i)^n D^n J^{2n}. \quad (3.28)$$

We then use these to form our deconstructed $\exp(\tfrac{1}{2}id\partial\delta^2)\exp(-\tfrac{1}{2}iDJ^2)$ of (3.8). Thus, we expand out each series, multiply these together, reorder terms in powers of *J*, and consolidate into a merged double series. The calculation is straightforward but tedious, and we show the first six orders, which are sufficient to establish the pattern for consolidation:

$$\exp\left(\tfrac{1}{2}id\,\partial\,\delta^2\right)\exp\left(-\tfrac{1}{2}i\,D\,J^2\right)=\operatorname{coshw}\left(\sqrt{i}\sqrt{d\partial}\,\delta\right)\operatorname{coshw}\left(i\sqrt{i}\sqrt{D}\,J\right)$$

$$=\left(\sum_{n=0}^{\infty}\frac{1}{2n!!}i^n(d\partial)^n\delta^{2n}\right)\left(\sum_{n=0}^{\infty}\frac{1}{2n!!}(-i)^n D^n J^{2n}\right)$$

$$=\left(1+\frac{1!!}{2!!}(d\partial D)+\frac{3!!}{4!!}(d\partial D)^2+\frac{5!!}{6!!}(d\partial D)^3+\frac{7!!}{8!!}(d\partial D)^4+\frac{9!!}{10!!}(d\partial D)^5+\frac{11!!}{12!!}(d\partial D)^6\right)\frac{1}{0!}D^0 J^0$$

$$-i\left(\frac{1!!}{0!!}+\frac{3!!}{2!!}(d\partial D)+\frac{5!!}{4!!}(d\partial D)^2+\frac{7!!}{6!!}(d\partial D)^3+\frac{9!!}{8!!}(d\partial D)^4+\frac{11!!}{10!!}(d\partial D)^5+\frac{13!!}{12!!}(d\partial D)^6\right)\frac{1}{2!}DJ^2$$

$$-\left(\frac{3!!}{0!!}+\frac{5!!}{2!!}(d\partial D)+\frac{7!!}{4!!}(d\partial D)^2+\frac{9!!}{6!!}(d\partial D)^3+\frac{11!!}{8!!}(d\partial D)^4+\frac{13!!}{10!!}(d\partial D)^5+\frac{15!!}{12!!}(d\partial D)^6\right)\frac{1}{4!}D^2 J^4$$

$$+i\left(\frac{5!!}{0!!}+\frac{7!!}{2!!}(d\partial D)+\frac{9!!}{4!!}(d\partial D)^2+\frac{11!!}{6!!}(d\partial D)^3+\frac{13!!}{8!!}(d\partial D)^4+\frac{15!!}{10!!}(d\partial D)^5+\frac{17!!}{12!!}(d\partial D)^6\right)\frac{1}{6!}D^3 J^6 \quad . \quad (3.29)$$

$$+\left(\frac{7!!}{0!!}+\frac{9!!}{2!!}(d\partial D)+\frac{11!!}{4!!}(d\partial D)^2+\frac{13!!}{6!!}(d\partial D)^3+\frac{15!!}{8!!}(d\partial D)^4+\frac{17!!}{10!!}(d\partial D)^5+\frac{19!!}{12!!}(d\partial D)^6\right)\frac{1}{8!}D^4 J^8$$

$$-i\left(\frac{9!!}{0!!}+\frac{11!!}{2!!}(d\partial D)+\frac{13!!}{4!!}(d\partial D)^2+\frac{15!!}{6!!}(d\partial D)^3+\frac{17!!}{8!!}(d\partial D)^4+\frac{19!!}{10!!}(d\partial D)^5+\frac{21!!}{12!!}(d\partial D)^6\right)\frac{1}{10!}D^5 J^{10}$$

$$-\left(\frac{11!!}{0!!}+\frac{13!!}{2!!}(d\partial D)+\frac{15!!}{4!!}(d\partial D)^2+\frac{17!!}{6!!}(d\partial D)^3+\frac{19!!}{8!!}(d\partial D)^4+\frac{21!!}{10!!}(d\partial D)^5+\frac{23!!}{12!!}(d\partial D)^6\right)\frac{1}{12!}D^6 J^{12}$$

$$=\sum_{n=0}^{\infty}\left(\sum_{m=0}^{\infty}(-i)^n\frac{(2n+2m-1)!!}{2m!!}(d\partial D)^m\right)\frac{1}{2n!}(DJ^2)^n$$

$$=\sum_{m=0}^{\infty}\left(\sum_{n=0}^{\infty}(-i)^n\frac{(2n+2m-1)!!}{2n!}D^n J^{2n}\right)\frac{1}{2m!!}(d\partial D)^m$$

Based on the final two lines above, we may write this expressly in terms of the Green functions $G^{(n)}$, or alternatively in terms of "Wick functions" $W^{(m)}$, as:

$$\exp\left(\tfrac{1}{2}id\,\partial\,\delta^2\right)\exp\left(-\tfrac{1}{2}i\,D\,J^2\right)=\operatorname{coshw}\left(\sqrt{i}\sqrt{d\partial}\,\delta\right)\operatorname{coshw}\left(i\sqrt{i}\sqrt{D}\,J\right)$$

$$=\sum_{n=0}^{\infty}G^{(n)}\frac{1}{2n!}(DJ^2)^n=\sum_{m=0}^{\infty}\frac{1}{2m!!}(d\partial D)^m W^{(m)}$$

- where -  . (3.30)

$$G^{(n)}\equiv\sum_{m=0}^{\infty}(-i)^n\frac{(2n+2m-1)!!}{2m!!}(d\partial D)^m$$

$$W^{(m)}\equiv\sum_{n=0}^{\infty}(-i)^n\frac{(2n+2m-1)!!}{2n!}(DJ^2)^n$$

The Green functions $G^{(n)}$ are thus specified as the expansion coefficients of $(1/2n!)DJ^2$ with a single factorial, while the Wick functions $W^{(m)}$ are the expansion coefficients of $(1/2m!!)(d\partial D)^m$ with a double factorial. The former Green expansion $\sum_{n=0}^{\infty} G^{(n)}(1/2n!)(DJ^2)^n$ mirrors the "ordinary" single-factorial Maclaurin series expansions for $f(x)$ in (3.26). The latter Wick expansion $\sum_{m=0}^{\infty}(1/2m!!)(d\partial D)^m W^{(m)}$ mirrors the parallel double-factorial "Wick-Maclaurin series" expansions for $fw(x)$ in (3.26). In fact, aside from the factor in (3.1) and (3.2), if we define the ordinary single factorial expansion function $f(DJ^2, G^{(n)}) \equiv \mathcal{C} \sum_{n=0}^{\infty} G^{(n)}(1/2n!)(DJ^2)^n$, and the corresponding "Wick-Maclaurin" double factorial expansion function $fw(d\partial D, W^{(m)}) \equiv \mathcal{C} \sum_{m=0}^{\infty} W^{(m)}(1/2m!!)(d\partial D)^m$, then (3.30) tell us, in terms of (3.26), that:

$$Z = f(DJ^2, G^{(n)}) = fw(d\partial D, W^{(m)}). \tag{3.31}$$

This is a concrete and concise application of Wick-Maclaurin functions to mathematically represent the statement (see [3], page 44) that the path integral can be expanded either in powers of sources $J$ (here, $f(DJ^2, G^{(n)})$), or in powers of "couplings" $\lambda$ (here, $fw(d\partial D, W^{(m)})$).

Finally, having found the "deconstructed" Green functions of (3.30), we turn now to "reconstruction." To do this, we need to "reverse" the deconstruction of (3.8) through (3.14). There are two terms in (3.30) which we need to reconstruct. The first, $DJ^2$, is simple. From (3.8), we can directly obtain the reconstruction:

$$DJ^2 \to \sqrt{-g} J^\mu D_{\mu\nu} \sqrt{-g} J^\nu. \tag{3.32}$$

The second term, $d\partial D$, is a little trickier, because it does not appear directly in (3.8). That, however, is why we developed (3.14) which is based on (3.6) and the "term of interest" (3.7). Specifically, combining (3.14) with (3.32) allows us to deduce that:

$$d\partial D \to \sqrt{-g} \partial_\mu \left( \partial^{:[\mu} D^{\{\nu]}{}_{\nu\}} \right). \tag{3.33}$$

Therefore, using (3.32) and (3.33) in (3.30), we can reconstruct all the way back to the local, non-integral equation (3.2), in the form of:

$$\exp iW(J(x)) = \sum_{n=0}^{\infty} G^{(n)} \frac{1}{2n!} \left( \sqrt{-g} J^\mu D_{\mu\nu} \sqrt{-g} J^\nu \right)^n = \sum_{m=0}^{\infty} \frac{1}{2m!!} \left( \sqrt{-g} \partial_\mu \left[ \partial^{:[\mu} D^{\{\nu]}{}_{\nu\}} \right] \right)^m W^{(m)}$$

- where -

$$G^{(n)} = \sum_{m=0}^{\infty} (-i)^n \frac{(2n+2m-1)!!}{2m!!} \left( \sqrt{-g} \partial_\mu \left[ \partial^{:[\mu} D^{\{\nu]}{}_{\nu\}} \right] \right)^m \tag{3.34}$$

$$W^{(m)} = \sum_{n=0}^{\infty} (-i)^n \frac{(2n+2m-1)!!}{2n!} \left( \sqrt{-g} J^\mu D_{\mu\nu} \sqrt{-g} J^\nu \right)^n$$

If we put all of this together explicitly, expanded in $J$, we have:

$$\exp iW(J(x)) = \sum_{n=0}^{\infty} \sum_{m=0}^{\infty} (-i)^n \frac{(2n+2m-1)!!}{2m!!} \left(\sqrt{-g}\partial_\mu \left[\partial^{:[\mu} D^{\{\nu]}{}_{\nu\}}\right]\right)^m \frac{1}{2n!} \left(\sqrt{-g}J^\mu D_{\mu\nu} \sqrt{-g}J^\nu\right)^n . \quad (3.35)$$

Now, reconstructing back to (3.1), we restore the non-locality and the integrals, and so finally arrive at the entire path integral $Z = \mathcal{C} \exp iW(J(x))$, in closed, exact form, as such:

$$\exp iW(J(x)) =$$
$$= \sum_{n=0}^{\infty} \sum_{m=0}^{\infty} (-i)^n \frac{(2n+2m-1)!!}{2m!!} \left(\int_U \sqrt{-g}d^4x \sqrt{-g}d^4y \partial_\mu \left[\partial^{:[\mu} D^{\{\nu]}{}_{\nu\}}(x-y)\right]\right)^m$$
$$\times \frac{1}{2n!} \left(\int_U \sqrt{-g}d^4x \sqrt{-g}d^4y J^\mu(x) D_{\mu\nu}(x-y) J^\nu(y)\right)^n \quad (3.36)$$

It should be noted that in the course of doing the expanded calculation (3.29), the operator term with $d\partial\delta^2$, which contains no explicit propagator, turned into terms expanded in $d\partial D$, that is, $d\partial\delta^2 \to d\partial D$ after functional differentiation, which *does* contain a propagator. When reconstructed and treated non-locally, this propagator becomes $D \to D^{\{\nu]}{}_{\nu\}}(x-y)$, which introduces the "two point" relationship $x-y$, which is why we need to also add the factor $\sqrt{-g}d^4x\sqrt{-g}d^4y$ to the $\partial_\mu\left[\partial^{:[\mu}D^{\{\nu]}{}_{\nu\}}(x-y)\right]$ term above.

Written in the form of (3.34), we may represent (3.36) as:

$$Z = \mathcal{C} \exp iW(J(x))$$
$$= \mathcal{C}\sum_{n=0}^{\infty} G^{(n)} \frac{1}{2n!} \left(\int_U \sqrt{-g}d^4x \sqrt{-g}d^4y J^\mu(x) D_{\mu\nu}(x-y) J^\nu(y)\right)^n$$
$$= \mathcal{C}\sum_{m=0}^{\infty} \frac{1}{2m!!} \left(\int_U \sqrt{-g}d^4x \sqrt{-g}d^4y \partial_\mu \left[\partial^{:[\mu} D^{\{\nu]}{}_{\nu\}}(x-y)\right]\right)^m W^{(m)} \quad (3.37)$$

- where -

$$G^{(n)} = \sum_{m=0}^{\infty} (-i)^n \frac{(2n+2m-1)!!}{2m!!} \left(\int_U \sqrt{-g}d^4x \sqrt{-g}d^4y \partial_\mu \left[\partial^{:[\mu} D^{\{\nu]}{}_{\nu\}}(x-y)\right]\right)^m$$

$$W^{(m)} = \sum_{n=0}^{\infty} (-i)^n \frac{(2n+2m-1)!!}{2n!} \left(\int_U \sqrt{-g}d^4x \sqrt{-g}d^4y J^\mu(x) D_{\mu\nu}(x-y) J^\nu(y)\right)^n$$

The final reconstruction step is to use Gauss / Stokes theorem $\int_U d^4x \partial_\mu \leftrightarrow \int_{\partial U} (d^3x)_\mu$ in the forward direction, and integrate (3.37) back to a boundary integral over $\partial U$. However, the pertinent term is $d^4x d^4y \partial_\mu$, because of the entry of the propagator $D \to D^{\{\nu]}{}_{\nu\}}(x-y)$ through $d\partial\delta^2 \to d\partial D$, as just discussed. In this situation, the appropriate Gauss / Stokes conversion is $\int_U d^4x d^4y \partial_\mu \leftrightarrow \int_{\partial U} (d^3x d^3y)_\mu$, where the two-point, three-volume element:

$$(d^3x d^3y)_\mu \equiv \tfrac{1}{3!}\varepsilon_{\mu\alpha\beta\sigma}(dxdy)^\alpha \wedge (dxdy)^\beta \wedge (dxdy)^\sigma. \tag{3.38}$$

Therefore, the final, complete, reconstructed QED path integral, in curved spacetime, in closed series form and including the integrated boundary term, is:

$$\boxed{\begin{aligned}
Z &= \mathcal{C}\exp iW(J(x)) \\
&= \mathcal{C}\sum_{n=0}^{\infty} G^{(n)} \frac{1}{2n!}\left(\int_U \sqrt{-g}d^4x\sqrt{-g}d^4y J^\mu(x) D_{\mu\nu}(x-y) J^\nu(y)\right)^n \\
&= \mathcal{C}\sum_{m=0}^{\infty} \frac{1}{2m!!}\left(\int_{\partial U}\left(\sqrt{-g}d^3x\sqrt{-g}d^3y\right)_\mu \partial^{[\mu} D^{\{\nu\}}{}_{\nu\}}(x-y)\right)^m W^{(m)} \\
&\text{where} \\
G^{(n)} &= \sum_{m=0}^{\infty}(-i)^n \frac{(2n+2m-1)!!}{2m!!}\left(\int_{\partial U}\left(\sqrt{-g}d^3x\sqrt{-g}d^3y\right)_\mu \partial^{[\mu} D^{\{\nu\}}{}_{\nu\}}(x-y)\right)^m \\
W^{(m)} &= \sum_{n=0}^{\infty}(-i)^n \frac{(2n+2m-1)!!}{2n!}\left(\int_U \sqrt{-g}d^4x\sqrt{-g}d^4y J^\mu(x) D_{\mu\nu}(x-y) J^\nu(y)\right)^n
\end{aligned}} \tag{3.39}$$

In the more abstract representation of (3.31), we may represent these alternative expansions as:

$$\boxed{\begin{aligned}
Z &= f\left(\int_U \sqrt{-g}d^4x\sqrt{-g}d^4y J^\mu(x) D_{\mu\nu}(x-y) J^\nu(y), G^{(n)}\right) \\
&= fw\left(\int_{\partial U}\left(\sqrt{-g}d^3x\sqrt{-g}d^3y\right)_\mu \partial^{[\mu} D^{\{\nu\}}{}_{\nu\}}(x-y), W^{(m)}\right)
\end{aligned}} \tag{3.40}$$

As is to be expected, each term in $W(J(x))$ may be parameterized by $m$ and $n$, that is, by $W_{(m,n)}(J(x))$. Now, let's take a closer look at the boundary term, by examining the amplitude $W_{(0,\sum n)}(J(x))$, that is, the amplitude terms for which $m=0$, and for which we take the sum over $n$ from 0 to $\infty$, represented by $\sum n$. In this circumstance, using the $G^{(n)}$ formulation, (3.39) becomes:

$$\begin{aligned}
&\exp iW_{(0,\sum n)}(J(x)) \\
&= \sum_{n=0}^{\infty} G^{(n)} \frac{1}{2n!}\left(\int_U \sqrt{-g}d^4x\sqrt{-g}d^4y J^\mu(x) D_{\mu\nu}(x-y) J^\nu(y)\right)^n \\
&\text{where} \\
&G^{(n)} = (-i)^n \frac{(2n-1)!!}{0!!}\left(\int_{\partial U}\left(\sqrt{-g}d^3x\sqrt{-g}d^3y\right)_\mu \partial^{[\mu} D^{\{\nu\}}{}_{\nu\}}(x-y)\right)^0 = (-i)^n (2n-1)!!
\end{aligned} \tag{3.41}$$

Using (3.24) and (3.25) written as $\exp-i\frac{1}{2}DJ^2 = \sum_{n=0}^{\infty}\frac{1}{2n!!}(-i)^n D^n J^{2n}$, and given that

$(2n-1)!!/2n! = 1/2n!!$, these therefore combine into:

$\exp iW_{(0,\sum n)}(J(x))$

$$= \sum_{n=0}^{\infty}(-i)^n \frac{1}{2n!!}\left(\int_U \sqrt{-g}d^4x\sqrt{-g}d^4y J^\mu(x)D_{\mu\nu}(x-y)J^\nu(y)\right)^n . \quad (3.42)$$

$$= \exp\left(-\tfrac{1}{2}i\int_U \sqrt{-g}d^4x\sqrt{-g}d^4y J^\mu(x)D_{\mu\nu}(x-y)J^\nu(y)\right)$$

More directly, this means that:

$$\boxed{W_{(0,\sum n)}(J(x)) = -\tfrac{1}{2}\int_U \sqrt{-g}d^4x\sqrt{-g}d^4y J^\mu(x)D_{\mu\nu}(x-y)J^\nu(y)}. \quad (3.43)$$

This is the usual QED amplitude function, albeit integrated over the $\sqrt{-g}d^4x\sqrt{-g}d^4y$ volume elements and so applicable in curved spacetime, and it serves as a check that (3.39) is a correct result (see [3] at eq. (I.5.4)). This also tell us that this usual amplitude, is the formulation in which the $m=0$ in the Green function $G^{(n)}$ and in which we employ $\sum n$ for $n$. Thus, the usual amplitude (3.43) uses a sum starting with the two-point Green function, and running all the way through the "infinite point" Green function. This also gives us the context in which to examine the boundary term, which comes into play for $m \neq 0$.

We saw in (1.12) that the boundary term $\int_{\partial U}(d^3x)_\mu A_\nu \partial^{[\mu}A^{\nu]}$ may be discarded by taking $A_\nu = 0$ over the boundary $\partial U$. This term becomes $\int_{\partial U}\left(\sqrt{-g}d^3x\sqrt{-g}d^3y\right)_\mu \partial^{;[\mu}D^{\{\nu\}}{}_{\nu\}}(x-y)$ following path integral quantization, and it is part of the Green function $G^{(n)}$ in (3.39). Using $g_{\mu\nu}D^{\mu\nu} = D^\mu{}_\mu = D_\nu{}^\nu$ and $\partial^{\mu;}D^\nu{}_\nu = \partial^\mu D^\nu{}_\nu$ for the trace term, the integrand in this boundary term expands, including the Christoffel connections, to:

$$\partial^{;[\mu}D^{\{\nu\}}{}_{\nu\}}(x-y) = g^{\mu\sigma}g_{\beta\nu}\partial_{;[\sigma}D^{\{\nu]\beta\}}(x-y)$$
$$= 2\partial^\mu D^\nu{}_\nu(x-y) - \partial^\nu D^{\{\mu\}}{}_{\nu\}}(x-y) + g^{\nu\sigma}\Gamma^\tau{}_{\sigma\nu}D^{\{\mu\}}{}_{\tau\}}(x-y) - \Gamma^\mu{}_{\sigma\tau}D^{\{\tau\sigma\}}(x-y) \quad (3.44)$$

For this term to become zero, we would have to take $D^{\tau\sigma}(x-y) = 0$ over the boundary $\partial U$. In the event $D^{\tau\sigma}(x-y) = 0$ over $\partial U$, all the $G^{(n)}$ terms would vanish for $m > 0$, and, as in (3.41), we would have, overall $G^{(n)} = (-i)^n(2n-1)!!$, leading to (3.43).

Therefore, (3.43) is the amplitude in the limit where the propagator $D^{\tau\sigma}(x-y)$ is taken to vanish over the boundary, $D^{\tau\sigma}(x-y)=0$ over $\partial U$ (in contrast to, and as a QED progression from, taking $A_\nu=0$ over the boundary as in (1.12)). In the event the propagator over the boundary is other than zero, the non-zero Green functions $G^{(n)}\neq 0$ for $m>0$ will cause the amplitude $W(J(x))$ to vary from (3.43). That is, any variation from (3.43), would have to come about because $D^{\tau\sigma}(x-y)\neq 0$ over $\partial U$.

Thus far, everything developed to this point has been based mathematical calculation carried out in spacetime. We have not at any point made any reference to momentum space or attempted to use any sort of Fourier transformation or more general type of harmonic analysis, given that we have been working with a curved spacetime background within which conventional Fourier methods of harmonic analysis are thought to be inapplicable. While we have successfully averted these questions dealing with harmonic analysis throughout the development of Green functions, as we turn to obtaining an explicit expression for the propagator $D_{\nu\alpha}(x-y)$ in curved spacetime, we can avoid this no longer. As noted at the end of section 2, we will now need to closely consider harmonic analysis and what happens to flat-spacetime Fourier analysis, in curved spacetime.

## 4. Harmonic Analysis in Curved Spacetime: Working Postulates and Corollaries

Harmonic analysis, which we shall loosely regard as the development of analogues to Fourier analysis in curved spacetime and more-generally in non-Euclidean spaces, is still an open area of research, with much that is not yet understood. What is well understood, however, is that the degree of symmetry of the spacetime manifold under consideration has a great deal to do with what type of harmonic analysis may be properly applied, and even whether any type of analysis can be applied whatsoever. The more restricted the symmetry, the more likely it is that one can effectively apply some form of harmonic analysis.

One symmetry restriction that we know must apply in a *physical,* spacetime manifold of any character, is *gauge symmetry*. As we shall now seek to demonstrate, when one is working in curved spacetime with gauge symmetry, the gauge symmetry greatly enhances one's ability to carry out Fourier-analog harmonic analysis, *even against this curved spacetime background.* The key aspect of gauge theory which enables this enhancement, is that *by definition*, a gauge theory is designed to maintain invariance with respect to transformations of the Fourier kernel of the form $\exp(-ip_\tau x^\tau) \to \exp(i\alpha(x))\exp(-ip_\tau x^\tau) = \exp[-i(p_\tau x^\tau - \alpha(x))]$. That is, the invariance of gauge theory under completely arbitrary transformations $p_\tau x^\tau \to p_\tau x^\tau - \alpha(x)$ of the exponent in the Fourier kernel, allows a modified form of traditional Fourier analysis to be employed even for some (but not all) types of curved spacetime manifolds, *so long as the theory under consideration is a gauge theory*. As we shall also see, this approach also introduces non-Abelian, i.e., Yang-Mills gauge theory, in a rather surprising way.

We begin our discussion by returning to equation (2.14), which is a *definition* of the propagator $D_{\nu\alpha}(x-y)$ as the inverse of the operator $g^{\mu\nu}(\partial_{;\sigma}\partial^{;\sigma} + \tfrac{1}{2}\kappa T + m^2) - \partial^{;\mu}\partial^{;\nu} - \kappa T^{\mu\nu}$, which we reproduce below:

$$\begin{aligned}\delta^\mu{}_\alpha \delta^{(4)}(x-y) &\equiv \left(g^{\mu\nu}(\partial_{;\sigma}\partial^{;\sigma} + m^2) - \partial^{;\nu}\partial^{;\mu}\right) D_{\nu\alpha}(x-y) \\ &= \left(g^{\mu\nu}(\partial_{;\sigma}\partial^{;\sigma} + m^2) - \partial^{;\mu}\partial^{;\nu} + R^{\mu\nu}\right) D_{\nu\alpha}(x-y) \\ &= \left(g^{\mu\nu}(\partial_{;\sigma}\partial^{;\sigma} + \tfrac{1}{2}\kappa T + m^2) - \partial^{;\mu}\partial^{;\nu} - \kappa T^{\mu\nu}\right) D_{\nu\alpha}(x-y)\end{aligned} \quad (4.1)$$

Certainly, the operator $g^{\mu\nu}(\partial_{;\sigma}\partial^{;\sigma} + \tfrac{1}{2}\kappa T + m^2) - \partial^{;\mu}\partial^{;\nu} - \kappa T^{\mu\nu}$ was carefully constructed to be consistent with non-commuting derivatives and parallel transport on a curved spacetime manifold and the presence of gravitational curvature and matter sources $T^{\mu\nu}$, and so is suitable for use in a curved manifold. By embedding the Einstein equation $-\kappa T^{\mu\nu} = R^{\mu\nu} - \tfrac{1}{2}g^{\mu\nu}R$, (4.1) is also imposing a constraint on the metric tensor, by requiring that the metric tensor $g_{\mu\nu}$ must be a solution of the Einstein equation for whatever matter source $T^{\mu\nu}$ one applies. Going back to the relationship between the symmetry of the manifold and the ability to apply harmonic analysis, one may think of (4.1) as already imposing some symmetry on the manifold, namely, those symmetry constraints which are imposed from on given matter source $T^{\mu\nu}$ on the metric tensor, via $-\kappa T^{\mu\nu} = R^{\mu\nu} - \tfrac{1}{2}g^{\mu\nu}R$. (A nearly-encyclopedic catalog of such solutions is provided by [4].) Because we wish for $D_{\nu\alpha}(x-y)$ to be the inverse of this operator, a Kronecker delta $\delta^\mu{}_\alpha$, which is a unit matrix, of Dirac deltas $\delta^{(4)}(x-y)$, which are unit perfect impulses, must be used to define the inverse. The questions about harmonic analysis begin to arise, however, from the presence of the Dirac delta $\delta^{(4)}(x-y)$ in (4.1), because this delta is the (inverse) Fourier transform of "1" and so may be *defined* using the *mathematical identity*:

$$\delta^{(4)}(x-y) \equiv \int \frac{d^4 p}{(2\pi)^4} e^{ip_\tau(x-y)^\tau} . \tag{4.2}$$

This contains the Fourier kernel $e^{ik_\tau(x-y)^\tau}$, and it is known that as a general rule, one cannot simply apply this kernel in curved spacetime without very careful consideration. But, as used in (4.2), and without factors of $\sqrt{-g}$ which we know must eventually permeate any integration done in curved spacetime, the kernel is completely appropriate, because (4.2) is a *mathematical identity* that exists irrespective of the type of geometric space which may be under consideration. That is, equation (4.2) is a *pre-geometric* equation which is no more and no less than a mathematical identity, and so it serves as something of an "anchor" in the complexities of curved spacetime. Therefore, we may appropriately combine (4.1) and (4.2) to write:

$$\begin{aligned}
\int \frac{d^4 p}{(2\pi)^4} \delta^\mu{}_\alpha e^{ip_\tau(x-y)^\tau} &\equiv \left( g^{\mu\nu}(\partial_{;\sigma}\partial^{;\sigma} + m^2) - \partial^{;\nu}\partial^{;\mu} \right) D_{\nu\alpha}(x-y) \\
&= \left( g^{\mu\nu}(\partial_{;\sigma}\partial^{;\sigma} + m^2) - \partial^{;\mu}\partial^{;\nu} + R^{\mu\nu} \right) D_{\nu\alpha}(x-y) \\
&= \left( g^{\mu\nu}(\partial_{;\sigma}\partial^{;\sigma} + \tfrac{1}{2}\kappa T + m^2) - \partial^{;\mu}\partial^{;\nu} - \kappa T^{\mu\nu} \right) D_{\nu\alpha}(x-y)
\end{aligned} \tag{4.3}$$

This is the place from which we can no longer avoid considering harmonic analysis in curved spacetime, because to make any further headway with (4.3), we must have some analog of the (inverse) Fourier transform for the propagator, that is, we need a reliable analog of $D_{\nu\alpha}(x-y) = \mathcal{F}^{-1} D_{\nu\alpha}(p)$. If we cannot develop such an analog, then we may be forced to abandon the entire notion that path integration can ever be applied in a fully calculable manner to curved spacetime. This, in turn, may well render a possible unification of quantum theory and gravitational theory impossible – at least via the path integration route – and would thereby undercut one of the two main pillars (the other being canonical quantization) upon which quantum filed theory is presently understood today, as the basis for eventual unification with gravitation. That is, the inability to extend some analog of Fourier transformation into curved spacetime would signal that path integration is OK for Euclidean spacetime but if we ever wish to consider quantization of non-Euclidean spacetime, then we need to scrap the path integral approach and find an entirely new basis for quantum theory in curved manifolds.[*] Given these

---

[*] Or, as noted in [3] at page 456, we must find some "embedding" mathematical operation which, in the flat spacetime limiting approximation, reduces to the path integral $\int D\varphi \exp(-(i/\hbar)\int d^4 x \mathcal{L}(\varphi))$. This ought not be

cascading, possibly devastating consequences of not being able to extend some form of Fourier-type analysis into curved spacetime – or at least into curved spacetime manifolds constrained by some degree of symmetry which does not exclude physically-realistic solutions – it behooves us to fully and vigorously explore the question of how to do Harmonic analysis – the analog of Fourier analysis – in non-Euclidean spacetime manifolds, *even if this means requiring the manifolds we consider to possess certain symmetries which permit this to be done*.

Lest one think that accepting symmetry restrictions on the manifolds we consider is a case of the "drunk looking for the quarter under the lamppost" because "that is where the light is," keep in mind that *physics*, as oppose to pure mathematics, is a study of the mathematics which reproduces and more or less explains what we observe *in the natural world*. As such, at the risk of stating the obvious, physics itself is fundamentally a "process of elimination" which of its very nature restricts out from consideration, many mathematical systems which may be completely internally-consistent as a logical matter, but which simply do not match up with observational data. So, given that only certain types of curved manifolds with certain symmetries can be subjected to consistent forms of harmonic analysis which enable us to retain the path integral formulation of quantum field theory, and by restricting ourselves to only those type of manifolds which admit some form of harmonic analysis, we *may* (not "are," but "may") in fact be restricting ourselves to the same types of manifolds which nature herself permits to exist, and eliminating from consideration manifolds which nature does not permit to exist.

That is, it is clear that while *mathematics* permits many different manifolds to exist, one should not assume that *physics* itself will permit any old mathematical manifold to exist in the observed natural world. Indeed, if physics should turn out to be essentially geometrodynamic in character, the whole point of physics would be to eliminate all but those manifolds which do exist or can exist in nature. Already discussed, for example, given a matter source $T^{\mu\nu}$, the Einstein equation $-\kappa T^{\mu\nu} = R^{\mu\nu} - \frac{1}{2}g^{\mu\nu}R$ which is embedded in (4.3) restricts out from consideration, any manifolds for which the metric tensor $g^{\mu\nu}$ is not a solution to this equation. Thus, in the union of gravitational and quantum field theory, it is to be expected that *on top of general relativity, quantum field theory will impose even further restrictions on the manifolds which are permitted in nature*, down to a very small, perhaps unique set of manifolds which are

---

eliminated as a possibility in principle. But, we shall not pursue this possibility here, but rather will push the path integral formalism as far as it can possibly go to determine its outer limits, if any.

permitted by, i.e., not excluded, by *both* general relativity and by quantum field theory.  Also, for example, one should have no problem at all with accepting the symmetry constraint that any physically-viable manifold must possess gauge symmetry, even though pure mathematics might consistently permit manifolds without such symmetry.  Similarly, one should have no problem with accepting the symmetry constraint that that any physically-viable manifold must be locally Minkowski, that is, that the tangent / orthonormal space at each event on the spacetime manifold must have the metric tensor $g_{\mu\nu} = \eta_{\mu\nu}$.  Nor should one have a problem eliminating from the universe of possibilities, a 27-diensional manifold, etc., in favor of a 4-dimensional manifold with one time and three space dimensions. Thus, the acceptance of certain symmetry constraints on the manifold may be, not a case of "compromising" on possibilities or "looking only under the light for the quarter," but a case of narrowing our consideration of "possibilities" down to only those "actualities" which nature herself permits, in the most fundamental spirit of physics research.

Given the foregoing, we now raise all of this to the level of two working postulates, and their immediate corollaries, to be developed to the point where these postulates can be disproven by a contradictory example, or cannot be disproven, as follows:

<u>*Working Postulate 1*</u>:  The path integral formalism $Z = \int D\varphi \exp\left(-(i/\hbar)\int d^4 x \mathcal{L}(\varphi)\right)$ remains a valid description of quantum reality, even in curved spacetime manifolds.

<u>*Working Postulate 2*</u>:  The only spacetime manifolds which are permitted by nature to exist in the physical world (as opposed to being possible in mathematics but not realized or "actualized" in nature), are those for which the closed form Green functions can be fully calculated from the path integral formalism.

These in turn lead to two corollaries:

<u>*Corollary 1*</u>:  Because the path integral formalism and the derivation of closed form Green functions cannot be extended into curved spacetime manifolds without engaging in some form of harmonic analysis analogous to Fourier transformations, *the only manifolds which nature permits to exist in the physical world, are those manifolds upon which one can perform some consistent, calculable form harmonic analysis.*  That's is, nature excludes from existing, any manifolds upon which one cannot calculate using some type of harmonic analysis.

*Corollary 2*: Because the ability to perform harmonic analysis on a manifold requires that the manifold must contain certain symmetries, *any and all manifolds which are permitted to exist in the physical world, must contain the symmetries which are necessary to permit harmonic analysis of those manifolds in the context of path integral quantization.*

By these postulates and corollaries, we actually use the path integral formulation as the basis for discarding from consideration, as "unphysical," any manifold for which path integration cannot be performed with some appropriate form harmonic analysis which is the analog of Fourier transformations, because that manifold does not possess the requisite symmetry to permit such analysis.

We also introduce the more-conventional postulate that all of the physics occurring in our manifold is gauge-invariant, that is, we require gauge symmetry. And, we postulate that tangent to each event in the manifold is am orthonormal Minkowski space with the metric tensor $\eta^{ab}$, which we express formally in terms of a vierbein $V^{\mu}{}_a(x)$ as: (See, e.g., [5], section 3.8.)

$$g^{\mu\nu} = V^{\mu}{}_a V^{\nu}{}_b \eta^{ab}. \tag{4.4}$$

Now, let us pinpoint some of the problems that one runs into in applying traditional Fourier analysis to curved spacetime, even for manifolds with a high degree of symmetry. *If* we were able to simply use the regular Fourier transform $\mathcal{F}$ in curved spacetime without having to make any changes – <u>which we cannot do</u> – then, for the propagator $D_{\nu\alpha}(x-y)$ in (4.3), we would use the inverse Fourier transform specified by:

$$\mathcal{F}^{-1} D_{\nu\alpha}(p) = D_{\nu\alpha}(x-y) = \int \frac{d^4 p}{(2\pi)^4} D_{\nu\alpha}(p) e^{ip_\tau (x-y)^\tau}. \tag{4.5}$$

In the above, $d^4 p = dp_0 dp_1 dp_2 dp_3$ is specified using lower-indexed volume elements over the momentum space, in contrast to $d^4 x = dx^0 dx^1 dx^2 dx^3$. We would then combine this with (4.3) and remove the integrands to obtain:

$$\begin{aligned}\delta^{\mu}{}_{\alpha} &\equiv \left(g^{\mu\nu}\left(\partial_{;\sigma}\partial^{;\sigma} + m^2\right) - \partial^{;\nu}\partial^{;\mu}\right) D_{\nu\alpha}(p) \\ &= \left(g^{\mu\nu}\left(\partial_{;\sigma}\partial^{;\sigma} + m^2\right) - \partial^{;\mu}\partial^{;\nu} + R^{\mu\nu}\right) D_{\nu\alpha}(p) \\ &= \left(g^{\mu\nu}\left(\partial_{;\sigma}\partial^{;\sigma} + \tfrac{1}{2}\kappa T + m^2\right) - \partial^{;\mu}\partial^{;\nu} - \kappa T^{\mu\nu}\right) D_{\nu\alpha}(p)\end{aligned} \tag{4.6}$$

Then, we would go about calculating this inversion in the usual way, but with the additional terms involving $R^{\mu\nu}$ and / or $T^{\mu\nu}$. But, the inverse transform in (4.5) has two problems: First, when we carry out the integral over $d^4p$, we are now talking about a momentum space being defined as the dual to a non-Euclidean, curved spacetime, so we expect there to be some expression involving $\sqrt{-g}$ inside of this integral. That does not yet appear in (4.5). Secondly, in curved spacetime – and even in a flat spacetime to which we apply a curvilinear coordinate transformation – the Fourier kernel $e^{ip_\tau(x-y)^\tau}$ does not have an invariant meaning, but changes even under general coordinate transformations in flat spacetime. Thus, this kernel most certainly fails as is, in a generally non-Euclidean manifold.

To resolve at least these two problems, let us now narrow our earlier postulates even further, by working with one particular type of harmonic analysis, namely, *Pontryagin duality*.[*] That is, we shall now consider only manifolds to which Pontryagin duality may be applied, and thus, manifolds which form a locally compact, topological, Abelian group. Pontryagin duality is but one type of harmonic analysis, and the manifolds to which it applies have the highest degree of symmetry. In order of decreasing symmetry, and without being limiting, one can also perform harmonic analysis on Lie groups, or on homogeneous spaces, or one can apply the theory of pseudodifferential operators and Fourier integral operators on a local basis. [6] In narrowing our consideration in this way to Pontryagin duality, we are selecting the simplest form of harmonic analysis for application to manifolds with the highest degree of symmetry, without excluding other techniques for harmonic analysis on manifolds with less symmetry, so long as those techniques satisfy our postulates that 1) they can be applied in the context of path integral quantization, and 2) they can be applied in such a way that one can calculate closed form Green functions from the path integral.

Therefore using Pontryagin duality as the framework for further development, we take the "original" space to be spacetime $x$, and the "dual" space to be four-dimensional momentum space $p$. Then, in relation to a function $G(x)$, we define the "forward" "Harmonic" ($\mathcal{H}$) transform $G(p) \equiv \mathcal{H}(G(x))$ and the "inverse" transform $G(x) \equiv \mathcal{H}^{-1}(G(p))$ as: [7]

---

[*] A. Neumaier has pointed out in dialogue on the physics newsgroup sci.physics.research that "as long as the space-time is diffeomorphic to a homogeneous space one can use a diffeomorphism to transform coordinates to that space, then do the Fourier analysis there, then transform back. . . . For a group representation approach, it is enough to have a homogenous space (still a highly symmetric space but less than a symmetric space), and there are infinitely many of these even in 4D (one just needs 4 independent Killing fields), some of them of high interest to cosmology. . . . On the other hand, the less symmetries there are the more difficult is the analysis, and only the symmetric space case is fully developed."

$$\begin{cases} G(p) \equiv \mathcal{H}(G(x)) = \int_{U(x)} d\mu(x) G(x) \overline{\chi(x,p)} \\ G(x) \equiv \mathcal{H}^{-1}(G(p)) = \int_{U(p)} d\nu(p) G(p) \chi(x,p) = \mathcal{H}^{-1}(\mathcal{H}(G(x))) \end{cases} \quad (4.7)$$

In (4.7), $d\mu(x)$ is the forward Haar measure on $x$, and $d\nu(p)$ is the inverse Haar measure on $p$. In relation to the usual "Fourier" ($\mathcal{F}$) analysis in flat spacetime, we have the correspondences $\mathcal{H}(G(x)) \to \mathcal{F}(G(x))$, $d\mu(x) \to d^4x$, and $d\nu(p) \to d^4p$. Further, we have $\overline{\chi(x,p)} \to e^{-ip_\sigma x^\sigma}$ and $\chi(x,p) \to e^{ip_\sigma x^\sigma}$, and so shall refer to $\chi(x,p)$ and its complex conjugate $\overline{\chi(x,p)}$ as the Fourier "kernel analogs." We shall normalize these such that $\overline{\chi(x,p)}\chi(x,p) = |\chi(x,p)|^2 = 1$. The overall relationship $G(x) \equiv \mathcal{H}^{-1}(\mathcal{H}(G(x)))$ tells us that the inverse Harmonic transform of a Harmonic transform of a function is *defined* to be identical to the original function.

Encapsulating all of the discussion in this section, we have now formally required that any the manifold we consider, in addition to having a metric tensor $g^{\mu\nu} = V^\mu{}_a V^\nu{}_b \eta^{ab}$ and having gauge symmetry, must be a manifold to which (4.7) may be properly applied. This "circular" definition, again, is akin to "looking under the light" for the quarter because "that is where the light is," insofar we are saying that we will use (4.7) on any and all manifolds for which we can use (4.7), and not on any others. But, again, this "circular" definition serves to eliminate certain manifolds from consideration – namely, those to which (4.7) cannot be applied – and, when extended to encompass any other methods of harmonic analysis which allow closed form development of Green functions from the path integral, may *perhaps* (not "certainly" but "perhaps") be a way of using quantum field theory to eliminate non-physical manifolds in favor of manifold physically-permitted by nature.

So, to simplify progress from here, and to avoid the need for a lengthy catalog of different types of symmetries of different types of manifolds (again, see [4]), we only consider manifolds upon which (4.7) can be used because those manifolds possess the requisite symmetries to permit (4.7) to be used. Thus, we will now use and develop (4.7) without further discussion of the particular types of manifolds to which (4.7) can be applied, other than to make central and fundamental use of gauge symmetry, which will be a crucial ingredient of developing (4.7) to the point where we can advance beyond equation (4.3), an do an exact calculation of the propagator even in curved spacetime.

## 5. Using Gauge Symmetry to Facilitate Harmonic Analysis in Curved Spacetime with Pontryagin duality

The Pontryagin duality equations (4.7) contain two crucial ingredients which it is now essential to develop mathematically. First, we must develop the Haar measures $d\mu(x)$ and $d\nu(p)$. Second, we must develop the "kernel analogs" $\overline{\chi(x,p)}$ and $\chi(x,p)$.

The "forward" Haar measure has a clear definition in curved spacetime, namely:

$$d\mu(x) \equiv \sqrt{-g(x)}d^4x, \tag{5.1}$$

which we know must be used in any integral over $d^4x$. This enables us to advance (4.7) to:

$$\begin{cases} G(p) \equiv \mathcal{H}(G(x)) = \int_{U(x)} \sqrt{-g(x)} d^4 x G(x) \overline{\chi(x,p)} \\ G(x) \equiv \mathcal{H}^{-1}(G(p)) = \int_{U(p)} d\nu(p) G(p) \chi(x,p) = \mathcal{H}^{-1}(\mathcal{H}(G(x))) \end{cases} \tag{5.2}$$

From here, however, we cannot obtain a definite expression for the "inverse" Haar measure $d\nu(p)$, without knowing the kernel analogs $\chi(x,p)$ and $\overline{\chi(x,p)}$ which, in flat spacetime, become $\overline{\chi(x,p)} \to e^{-ip_\sigma x^\sigma}$ and $\chi(x,p) \to e^{ip_\sigma x^\sigma}$. This brings us finally to gauge theory, where as observed at the start of section 4, the factor $e^{-ip_\sigma x^\sigma}$ is also a central element.

We start our discussion of gauge theory simply with a free fermion wavefunction and adjoint specified in the usual way:

$$\psi = e^{-ip_\sigma x^\sigma} u(p); \quad \overline{\psi} = \psi^\dagger \gamma^0 = e^{ip_\sigma x^\sigma} u^\dagger(p) \gamma^0 = e^{ip_\sigma x^\sigma} \overline{u}(p), \tag{5.3}$$

in flat spacetime in rectilinear coordinates. These wavefunctions are clearly specified using the Fourier kernel $e^{-ip_\sigma x^\sigma}$. When used in the Dirac Lagrangian density, we have:

$$\mathcal{L} = i\overline{\psi}\gamma^\mu \partial_\mu \psi - m\overline{\psi}\psi = \overline{\psi}\gamma^\mu p_\mu \psi - m\overline{\psi}\psi. \tag{5.4}$$

Gauge theory begins by subjecting these wavefunctions to a local gauge transformation:

$$\begin{aligned} \psi &= e^{-ip_\sigma x^\sigma} u(p) \to \psi' = e^{i\alpha(x)} \psi = e^{-ip_\sigma x^\sigma} e^{i\alpha(x)} u(p) = e^{-i(p_\sigma x^\sigma - \alpha(x))} u(p) \\ \overline{\psi} &= e^{ip_\sigma x^\sigma} \overline{u}(p) \to = \overline{\psi}' = e^{-i\alpha(x)} \overline{\psi} = e^{ip_\sigma x^\sigma} e^{-i\alpha(x)} \overline{u}(p) = e^{i(p_\sigma x^\sigma - \alpha(x))} \overline{u}(p) \end{aligned}. \tag{5.5}$$

In the Lagrangian density, this results in:

$$\mathcal{L} \to \mathcal{L}' = i\overline{\psi}'\gamma^\mu \partial_\mu \psi' - m\overline{\psi}'\psi' = \overline{\psi}\gamma^\mu (p_\mu + i\partial_\mu \alpha)\psi - m\overline{\psi}\psi. \tag{5.6}$$

This, of course, does not have the same form as (5.4). We therefore impose the requirement that $\mathcal{L}$ must be invariant under such local gauge transformations, and so introduce a gauge field $A_\mu$ and charge strength $e$ and use these to construct a "covariant" derivative:

$$D_\mu \equiv \partial_\mu - ieA_\mu. \tag{5.7}$$

We then use (5.7) to redefine the Lagrangian density, according to:

$$\mathcal{L} = i\bar{\psi}\gamma^\mu D_\mu \psi - m\bar{\psi}\psi = \bar{\psi}\gamma^\mu(i\partial_\mu + eA_\mu)\psi - m\bar{\psi}\psi = \bar{\psi}\gamma^\mu(p_\mu + eA_\mu)\psi - m\bar{\psi}\psi. \tag{5.8}$$

Now, under a gauge transformation (5.5), we have:

$$\mathcal{L} \to \mathcal{L}' = i\bar{\psi}\gamma^\mu(\partial_\mu + ieA_\mu + i\partial_\mu \alpha)\psi - m\bar{\psi}\psi \equiv i\bar{\psi}\gamma^\mu(\partial_\mu + ieA'_\mu)\psi - m\bar{\psi}\psi, \tag{5.9}$$

where, in the last expression, we have defined the transformed gauge field:

$$A_\mu \to A'_\mu = A_\mu + \frac{1}{e}\partial_\mu \alpha. \tag{5.10}$$

Once we interpret $A_\mu$ as the physical vector potential / photon field appearing in equation (1.1), then we can develop the continuity equation $\partial_\mu J^\mu = 0$ to identify the current density $J^\mu = e\bar{\psi}\gamma^\mu\psi$, and we must add to the Lagrangian (5.8), the term $-\frac{1}{4}F_{\mu\nu}F^{\mu\nu}$ for the kinetic energy of this gauge field, leading to: (Essentially the same development as the foregoing is presented in section 14.3 of [8].)

$$\mathcal{L} = i\bar{\psi}\gamma^\mu \partial_\mu \psi - \frac{1}{4}F_{\mu\nu}F^{\mu\nu} + e\bar{\psi}\gamma^\mu A_\mu \psi - m\bar{\psi}\psi = i\bar{\psi}\gamma^\mu \partial_\mu \psi - m\bar{\psi}\psi - \frac{1}{4}F_{\mu\nu}F^{\mu\nu} + A_\mu J^\mu. \tag{5.11}$$

This brings us full circle back to (1.2) from which we obtained the integration-by-parts, as well as (1.10), in which we first added the source term $A_\mu J^\mu$.

But, the lesson of repeating this well-known exercise in the development of gauge theory is one that works in reverse: Suppose we start out with the wavefunction $\psi' = e^{-i(p_\sigma x^\sigma - \alpha(x))}u(p)$ from (5.5), together with the corresponding photon field $A'_\mu = A_\mu + (1/e)\partial_\mu \alpha$ of (5.10). If we wish, the gauge invariance of our theory allows us to transform $A'_\mu \to A_\mu = A'_\mu - (1/e)\partial_\mu \alpha$ in the reverse direction, and in so doing, we will simultaneously transform in the reverse direction, $\psi' = e^{-i(p_\sigma x^\sigma - \alpha(x))}u(p) \to \psi = e^{-ip_\sigma x^\sigma}u(p)$. That is, we can "gauge out" the phase $\alpha(x)$ at will, and in particular, re-gauge the Fourier kernel $\bar{K}$ according to:

$$\bar{K}' = e^{-i(p_\sigma x^\sigma - \alpha(x))} \to \bar{K} = e^{-ip_\sigma x^\sigma}, \tag{5.12}$$

at will, without in any way without in any way affecting the gauge invariance of our theory.

Now, introducing $\theta(x,p)$, which taken to be a completely arbitrary, local function of spacetime and momentum space, and recognizing that $\overline{K} = e^{-ip_\sigma x^\sigma}$ is also the Fourier kernel in flat spacetime, we wish to consider the possibility that $\overline{\chi(x,p)} \equiv \overline{K'} = e^{-i(p_\sigma x^\sigma - \theta(x,p))}$ might be a suitable candidate for the kernel analog in curved spacetime for a manifold to which Pontryagin duality may be applied, and whether, by a suitable re-gauging of $\overline{\chi(x,p)} \to \overline{\chi'(x,p)} = e^{-ip_\sigma x^\sigma}$, just as in (5.12), it might be possible to employ the ordinary Fourier kernel $e^{-ip_\sigma x^\sigma}$, even in curved spacetime, without in any way sacrificing the invariance of our theory, precisely because gauge symmetry allows us to re-gauge a generalized kernel of the form $e^{-i(p_\sigma x^\sigma - \alpha(x))}$ into the linear kernel $e^{-ip_\sigma x^\sigma}$.

As a next step in our exploration, let us consider a Fourier kernel $e^{-ip_\sigma x^\sigma}$ in rectilinear coordinates. Keeping in mind the very important point that the coordinates $x^\mu$ are themselves *not a vector* under general coordinate transformations, let us then transform this Kernel into some arbitrary system of curvilinear coordinates:

$$x^\nu \to x'^\nu = x^\nu - \Lambda^\nu(x), \tag{5.13}$$

where $\Lambda^\nu(x)$ is a four-component, quadruplet, *local* parameter which varies as a generalized function of the spacetime coordinates $x^\nu$, based on the chosen, arbitrary system of coordinates. The momentum $p_\mu$, of course, is a vector, and so transforms as:

$$p_\mu \to p'_\mu = \frac{\partial x^\sigma}{\partial x'^\mu} p_\sigma. \tag{5.14}$$

Similarly, the partial derivative transforms as:

$$\partial_\mu \to \partial'_\mu = \frac{\partial x^\tau}{\partial x'^\mu} \partial_\tau. \tag{5.15}$$

From (5.13) and (5.15), we then deduce the intermediate result: (see just before equation (VIII.I.8) in [3])

$$\partial'_\mu x^\sigma = \frac{\partial x^\sigma}{\partial x'^\mu} = \frac{\partial x'^\sigma}{\partial x'^\mu} + \frac{\partial \Lambda^\sigma}{\partial x'^\mu} = \delta^\sigma{}_\mu + \partial'_\mu \Lambda^\sigma, \tag{5.16}$$

which then allows us to determine from all of (5.13) through (5.16) that under a general transformation:

$$p_\mu x^\mu \rightarrow p'_\mu x'^\mu = \frac{\partial x^\sigma}{\partial x'^\mu} p_\sigma (x^\mu - \Lambda^\mu(x^\nu)) = p_\sigma(\delta^\sigma{}_\mu + \partial'_\mu \Lambda^\sigma)(x^\mu - \Lambda^\mu(x^\nu))$$
$$= p_\mu x^\mu - p_\mu \Lambda^\mu(x^\nu) + p_\mu \partial'_\sigma \Lambda^\mu(x^\sigma - \Lambda^\sigma(x^\nu)) \quad , \quad (5.17)$$
$$\equiv p_\mu x^\mu - \beta(p, x)$$

where in the final line we have *defined* a singlet local parameter which depends directly upon the chosen coordinate transformation:

$$\beta(p, x) \equiv p_\mu \Lambda^\mu(x^\nu) - p_\mu \partial'_\sigma \Lambda^\mu(x^\sigma - \Lambda^\sigma(x^\nu)) = p_\mu x^\mu - p'_\mu x'^\mu . \quad (5.18)$$

Therefore, under this same transformation, the Fourier kernel transforms as:

$$e^{-ip_\mu x^\mu} \rightarrow e^{-ip'_\mu x'^\mu} = e^{-i(p_\mu x^\mu - \beta(p,x))} = e^{i\beta(p,x)} e^{-ip_\mu x^\mu} . \quad (5.19)$$

Suppose, therefore, that we now chose to define the Fourier kernel analog in the context of Pontryagin duality, in rectilinear coordinates, according to:

$$\overline{\chi(x, p)} \equiv e^{-i(p_\sigma x^\sigma - \theta(x,p))} = e^{i\theta(x,p)} e^{-ip_\sigma x^\sigma} , \quad (5.20)$$

where $\theta(x, p)$ is a completely arbitrary, local function of spacetime and momentum space, as already stated. Suppose further that we then perform general coordinate transformation on this kernel, so that we now have:

$$\overline{\chi(x, p)} \rightarrow \overline{\chi'(x, p)} = e^{i\beta(p,x)} \overline{\chi(x, p)} = e^{i\beta(p,x)} e^{i\theta(x,p)} e^{-ip_\sigma x^\sigma} , \quad (5.21)$$

with $\beta(p, x)$ as deduced in (5.18). Now, let's return to (5.5). The phase parameter $\alpha(x)$ in (5.5) is a completely arbitrary, local function of spacetime. However, there is nothing which prevents us from also selecting this to be an arbitrary function of momentum space as well, $\alpha(x) \rightarrow \alpha(x, p)$. After all, $\partial_\mu e^{i\alpha(x)} = i\partial_\mu \alpha(x)$ and $\partial_\mu e^{i\alpha(x,p)} = i\partial_\mu \alpha(x, p)$ have precisely the same form since $\partial_\mu$ does not operate on $p$, so that this extension to momentum space does nothing to compromise the gauge symmetry of our theory. Therefore, on top of the arbitrariness of $\theta(x, p)$ and the arbitrariness of the coordinate transformation (5.13) leading to another arbitrary $\beta(p, x)$ in (5.18), let us also perform an arbitrary gauge transformation on the kernel analog in (5.21). Now, we have:

$$\overline{\chi(x, p)} \rightarrow \overline{\chi'(x, p)} \rightarrow \overline{\chi''(x, p)} = e^{i\alpha(p,x)} \overline{\chi'(x, p)} = e^{i\alpha(p,x)} e^{i\beta(p,x)} e^{i\theta(x,p)} e^{-ip_\sigma x^\sigma} , \quad (5.22)$$

But, the gauge parameter $\alpha(x,p)$ is completely arbitrary, local function of $x$ and $p$, so we can select this parameter however we wish. So, we select $\alpha(x,p)$ such that:

$$\alpha(p,x) = -\beta(p,x) - \theta(x,p), \tag{5.23}$$

which means that

$$\overline{\chi'(x,p)} = e^{-ip_\sigma x^\sigma}, \tag{5.24}$$

which is the usual Fourier kernel. All that happens, is that the gauge field has now been transformed according to $A_\mu \to = A_\mu + (1/e)\partial_\mu \alpha$, with $\alpha(p,x)$ chosen according to (5.23). Thus, we have gauged out the general arbitrariness in (5.20), we have gauged out the further arbitrariness of the general coordinate transformation as expressed in (5.21), and we have returned the kernel to the Fourier kernel shown in (5.24). Because the gauge symmetry is local, we can do this at each and every event on the manifold.

As such, though we can select $\overline{\chi(x,p)} \equiv e^{-i(p_\sigma x^\sigma - \theta(x,p))}$ to contain the completely arbitrary local parameter $\theta(x,p)$, and though we can perform any arbitrary local coordinate transformation we wish, in the end, we can always gauge away all of this arbitrariness by selecting a local gauge transformation that allows us to transform to $\overline{\chi(x,p)} \to e^{-ip_\sigma x^\sigma}$ at each and every even on the manifold. Because of this, we might as well select the simplest form, and use:

$$\overline{\chi(x,p)} \equiv e^{-ip_\sigma x^\sigma}; \quad \chi(x,p) = e^{ip_\sigma x^\sigma}, \tag{5.25}$$

at each and every event, even in curved spacetime, and even if we are using general coordinates, because not matter how arbitrary $\chi(x,p)$ may become as a local function of spacetime and no matter how arbitrary our coordinate choice, *we can always find an arbitrary local gauge transformation that allows us to use (5.25) everywhere on the manifold, so long as the manifold is suitable for Pontryagin duality*.

Thus, we return to (5.2), and using (5.25), we now write:

$$\begin{cases} G(p) = \mathcal{H}(G(x)) = \int_{U(x)} \sqrt{-g(x)} d^4x\, G(x) e^{-ip_\sigma x^\sigma} \\ G(x) = \mathcal{H}^{-1}(G(p)) = \int_{U(p)} dv(p) G(p) e^{ip_\sigma x^\sigma} = \mathcal{H}^{-1}(\mathcal{H}(G(x))) \end{cases}. \tag{5.26}$$

Again, by virtue of gauge symmetry, we can always make a local choice of gauge which permits these relationships to be true everywhere in spacetime and in momentum space. Now, all that remains is to deduce the inverse Haar measure $dv(p)$, which we now write as:

$$dv(p) \equiv \frac{1}{(2\pi)^4} \omega d^4 p, \tag{5.27}$$

with $\omega$ an unknown expression to now be determined.

Because the usual Fourier kernels $e^{-ip_\sigma x^\sigma}$ and $e^{ip_\sigma x^\sigma}$ are now part of (5.26), we see that the harmonic transform in curved spacetime entails the *convolution* ("*") of the metric tensor determinant factor $\sqrt{-g(x)}$ with the subject function $G(x)$. Thus, from (5.26), we may write the forward Harmonic transform as:

$$G(p) = \mathcal{H}(G(x)) = \int_{U(x)} \sqrt{-g(x)} d^4 x G(x) e^{-ip_\sigma x^\sigma} = \mathcal{F}\left(\sqrt{-g(x)} G(x)\right) = \mathcal{F}\left(\sqrt{-g(x)}\right) * \mathcal{F}(G(x)), \tag{5.28}$$

whereby $\mathcal{H}(G(x)) = \mathcal{F}\left(\sqrt{-g(x)}\right) * \mathcal{F}(G(x))$ precisely expresses the harmonic transform $\mathcal{H}(G(x))$ as a convolution of the separate Fourier transforms of $\sqrt{-g(x)}$ and $G(x)$. Then, substituting this as well as (5.27) into the inverse transform in (5.26), using $G(x) = \mathcal{H}^{-1}(\mathcal{H}(G(x)))$, we obtain:

$$\begin{aligned} G(x) = \mathcal{H}^{-1}(G(p)) &= \int_{U(p)} \omega d^4 p G(p) e^{ip_\sigma x^\sigma} = \int_{U(p)} \omega d^4 p \mathcal{F}\left(\sqrt{-g(x)}\right) * \mathcal{F}(G(x)) e^{ip_\sigma x^\sigma} \\ &= \mathcal{F}^{-1}\left(\omega \cdot \mathcal{F}\left(\sqrt{-g(x)}\right) * \mathcal{F}(G(x))\right) \\ &= \mathcal{F}^{-1}(\omega) * \mathcal{F}^{-1}\left(\mathcal{F}\left(\sqrt{-g(x)}\right) * \mathcal{F}(G(x))\right) \\ &= \mathcal{F}^{-1}(\omega) * \sqrt{-g(x)} \cdot G(x) \end{aligned} \tag{5.29}$$

Now, we can factor out the $G(x)$, to find that:

$$1 = \mathcal{F}^{-1}(\omega) * \sqrt{-g(x)}. \tag{5.30}$$

In the above, $1 = 1(x)$. Fourier transforming one last time finally tells us that:

$$\delta^{(4)}(p) = \mathcal{F}(1) = \mathcal{F}\left(\mathcal{F}^{-1}(\omega) * \sqrt{-g(x)}\right) = \mathcal{F}\left(\mathcal{F}^{-1}(\omega)\right) \cdot \mathcal{F}\left(\sqrt{-g(x)}\right) = \omega \cdot \mathcal{F}\left(\sqrt{-g(x)}\right), \tag{5.31}$$

or:

$$\omega = \frac{1}{\mathcal{F}\left(\sqrt{-g(x)}\right)} \delta^{(4)}(p). \tag{5.32}$$

From (5.27), this means that the inverse Haar measure:

$$dv(p) \equiv \frac{1}{(2\pi)^4} \omega d^4 p = \frac{1}{(2\pi)^4} \frac{\delta^{(4)}(p)}{\mathcal{F}\left(\sqrt{-g(x)}\right)} d^4 p = \frac{1}{(2\pi)^4} \frac{\int d^4 x e^{-ip_\sigma x^\sigma}}{\int d^4 x \sqrt{-g(x)} e^{-ip_\sigma x^\sigma}} d^4 p. \tag{5.33}$$

Thus, from (5.26), using (5.33), we finally obtain the forward and inverse transforms:

$$\begin{cases} G(p) = \mathcal{H}(G(x)) = \int_{U(x)} \sqrt{-g(x)}\, d^4 x\, G(x) e^{-ip_\sigma x^\sigma} \\ G(x) = \mathcal{H}^{-1}(G(p)) = \int_{U(p)} \frac{1}{(2\pi)^4} \frac{\delta^{(4)}(p)}{\mathcal{F}\left(\sqrt{-g(x)}\right)} d^4 p\, G(p) e^{ip_\sigma x^\sigma} \end{cases}. \tag{5.34}$$

Finally, at very long last, we may return to (4.3), and complete our specification of the propagator inverse. Using the inverse transform in (5.34), for $G(x) = D_{\nu\alpha}(x-y)$, we may write:

$$D_{\nu\alpha}(x-y) = \mathcal{H}^{-1}(D_{\nu\alpha}(p)) = \int_{U(p)} \frac{1}{(2\pi)^4} \frac{\delta^{(4)}(p)}{\mathcal{F}\left(\sqrt{-g(x)}\right)} d^4 p\, D_{\nu\alpha}(p) e^{ip_\sigma (x-y)^\sigma}. \tag{5.35}$$

Then, using this in (4.3), removing the integrand and reducing, allows us, finally, to write:

$$\begin{aligned} \mathcal{F}\left(\sqrt{-g(x)}\right)\delta^\mu{}_\alpha &= \delta^{(4)}(p)\left(g^{\mu\nu}\left(\partial_{;\sigma}\partial^{;\sigma} + m^2\right) - \partial^{;\nu}\partial^{;\mu}\right) D_{\nu\alpha}(p) \\ &= \delta^{(4)}(p)\left(g^{\mu\nu}\left(\partial_{;\sigma}\partial^{;\sigma} + m^2\right) - \partial^{;\mu}\partial^{;\nu} + R^{\mu\nu}\right) D_{\nu\alpha}(p) \\ &= \delta^{(4)}(p)\left(g^{\mu\nu}\left(\partial_{;\sigma}\partial^{;\sigma} + \tfrac{1}{2}\kappa T + m^2\right) - \partial^{;\mu}\partial^{;\nu} - \kappa T^{\mu\nu}\right) D_{\nu\alpha}(p) \end{aligned}. \tag{5.36}$$

As we approach flat spacetime, $\sqrt{-g(x)} \to 1$, so $\mathcal{F}\left(\sqrt{-g(x)}\right) \to \delta^{(4)}(p)$, and the $\delta^{(4)}(p)$ cancel from each side. Further, if $R^{\mu\nu} = T^{\mu\nu} = 0$, in which case $\partial^{;\nu}\partial^{;\mu} \to \partial^\nu \partial^\mu$ and we commute derivatives, this reduces to the usual, familiar:

$$\delta^\mu{}_\alpha = \left(g^{\mu\nu}\left(\partial_\sigma \partial^\sigma + m^2\right) - \partial^\mu \partial^\nu\right) D_{\nu\alpha}(p) \tag{5.37}$$

This serves as a check, that all of the above does yield the correct result in the known, flat spacetime limit.

## 6. An Important Subtlety Giving Rise to Non-Abelian, Yang-Mills Gauge Theory

Let us now seek to solve (5.36). First, we use $\delta^{(4)}(p) = \mathcal{F}(1) = \int d^4 x\, e^{-ip_\tau x^\tau}$ for the Dirac impulse in (5.36), to write:

$$\begin{aligned} \mathcal{F}\left(\sqrt{-g(x)}\right)\delta^\mu{}_\alpha &= \left(g^{\mu\nu}\left(\partial_{;\sigma}\partial^{;\sigma} + m^2\right) - \partial^{;\nu}\partial^{;\mu}\right)\int d^4 x\, e^{-ip_\tau x^\tau} D_{\nu\alpha}(p) \\ &= \left(g^{\mu\nu}\left(\partial_{;\sigma}\partial^{;\sigma} + m^2\right) - \partial^{;\mu}\partial^{;\nu} + R^{\mu\nu}\right)\int d^4 x\, e^{-ip_\tau x^\tau} D_{\nu\alpha}(p) \\ &= \left(g^{\mu\nu}\left(\partial_{;\sigma}\partial^{;\sigma} + \tfrac{1}{2}\kappa T + m^2\right) - \partial^{;\mu}\partial^{;\nu} - \kappa T^{\mu\nu}\right)\int d^4 x\, e^{-ip_\tau x^\tau} D_{\nu\alpha}(p) \end{aligned}. \tag{6.1}$$

Now, while it is tempting to plow forward and use the various $\partial^{;\nu}$ in the above to operate on $e^{-ip_\tau x^\tau}$ and so make the substitution $\partial^{;\nu} e^{-ip_\tau x^\tau} \to -ip^\nu e^{-ip_\tau x^\tau}$ throughout, there is an important subtlety at work on (6.1) which arises because we are working in curved spacetime and because we have taken advantage of gauge symmetry in the previous section in order to be able render (4.3) calculable, leading to (5.36) and (6.1). And, this important subtlety does no more and no less than give rise to non-Abelian, Yang-Mills gauge theory, from a totally different approach than the usual.

Specifically, before we do anything further, let us subject the kernel in (6.1) to an arbitrary gauge transformation of the usual form $e^{-ip_\tau x^\tau} \to e^{i\alpha(p,x)} e^{-ip_\tau x^\tau} = e^{-i(p_\tau x^\tau - \alpha(p,x))}$. Then, if we operate with the various partial derivatives, (6.1) changes its form, to:

$$\mathcal{F}\left(\sqrt{-g(x)}\right)\delta^\mu{}_\alpha$$
$$= \left(g^{\mu\nu}\left(-(p_\sigma - \partial_\sigma \alpha)(p^\sigma - \partial^\sigma \alpha) + m^2\right) + (p^\nu - \partial^\nu \alpha)(p^\mu - \partial^\mu \alpha)\right)\int d^4 x e^{-i(p_\tau x^\tau - \alpha(p,x))} D_{\nu\alpha}(p)$$
$$= \left(g^{\mu\nu}\left(-(p_\sigma - \partial_\sigma \alpha)(p^\sigma - \partial^\sigma \alpha) + m^2\right) + (p^\mu - \partial^\mu \alpha)(p^\nu - \partial^\nu \alpha) + R^{\mu\nu}\right)\int d^4 x e^{-i(p_\tau x^\tau - \alpha(p,x))} D_{\nu\alpha}(p) \quad (6.2)$$
$$= \left(g^{\mu\nu}\left(-(p_\sigma - \partial_\sigma \alpha)(p^\sigma - \partial^\sigma \alpha) + \tfrac{1}{2}\kappa T + m^2\right) + (p^\mu - \partial^\mu \alpha)(p^\nu - \partial^\nu \alpha) - \kappa T^{\mu\nu}\right)\int d^4 x e^{-i(p_\tau x^\tau - \alpha(p,x))} D_{\nu\alpha}(p)$$

In the above, we have $\partial_{;\sigma} e^{-i(p_\tau x^\tau - \alpha(p,x))} = -i(p_\sigma - \partial_\sigma \alpha) e^{-i(p_\tau x^\tau - \alpha(p,x))}$ permitting the heuristic substitution $\partial_{;\sigma} \to -i(p_\sigma - \partial_\sigma \alpha)$, where we can take $\partial_{;\sigma} x^\tau = \partial_\sigma x^\tau = \delta^\tau{}_\nu$ because each of the four coordinates is a separate scalar field for which the covariant derivative is equal to the partial derivative, and because $\partial_{;\sigma} \alpha = \partial_\sigma \alpha$. This is exactly what happened in (5.6) which required us to introduce the partial derivative (5.7), and here too, we need to do the exact same thing. That is, to ensure that (6.1) remains invariant under local gauge transformations, so that we are in a position to "gauge out" any arbitrariness in the Pontryagin kernel $\overline{\chi(x,p)}$ of (5.20), and to "gauge out" any coordinate transformations $\overline{\chi(x,p)} \to \overline{\chi'(x,p)}$ of (5.21) on the kernel, we must replace each of the derivatives in (6.1) with the gauge-covariant derivative (5.7), generalized via the minimal coupling principle to:

$$D_{;\mu} \equiv \partial_{;\mu} - ieA_\mu. \tag{6.3}$$

Therefore, (6.1) must now become:

$$\mathcal{F}\left(\sqrt{-g(x)}\right)\delta^\mu{}_\alpha = \left(g^{\mu\nu}\left(D_{;\sigma} D^{;\sigma} + m^2\right) - D^{;\nu} D^{;\mu}\right)\int d^4 x e^{-ip_\tau x^\tau} D_{\nu\alpha}(p)$$
$$= \left(g^{\mu\nu}\left(D_{;\sigma} D^{;\sigma} + m^2\right) - D^{;\mu} D^{;\nu} + R^{\mu\nu}\right)\int d^4 x e^{-ip_\tau x^\tau} D_{\nu\alpha}(p) \quad (6.4)$$
$$= \left(g^{\mu\nu}\left(D_{;\sigma} D^{;\sigma} + \tfrac{1}{2}\kappa T + m^2\right) - D^{;\mu} D^{;\nu} - \kappa T^{\mu\nu}\right)\int d^4 x e^{-ip_\tau x^\tau} D_{\nu\alpha}(p)$$

Expanded out, and with $\partial_{;\sigma} e^{-ip_\tau x^\tau} \to -ip_\sigma e^{-ip_\tau x^\tau}$, this now reads:

$$\begin{aligned}
&\mathcal{F}\left(\sqrt{-g(x)}\right)\delta^{\mu}{}_{\alpha}\\
&=\left(g^{\mu\nu}\left(-(p_{\sigma}+eA_{\sigma})(p^{\sigma}+ieA^{\sigma})+m^{2}\right)+(p^{\nu}+eA^{\nu})(p^{\mu}+eA^{\mu})\right)\int d^{4}xe^{-ip_{\tau}x^{\tau}}D_{\nu\alpha}(p)\\
&=\left(g^{\mu\nu}\left(-(p_{\sigma}+eA_{\sigma})(p^{\sigma}+ieA^{\sigma})+m^{2}\right)+(p^{\mu}+eA^{\mu})(p^{\nu}+eA^{\nu})+R^{\mu\nu}\right)\int d^{4}xe^{-ip_{\tau}x^{\tau}}D_{\nu\alpha}(p)\\
&=\left(g^{\mu\nu}\left(-(p_{\sigma}+eA_{\sigma})(p^{\sigma}+ieA^{\sigma})+\tfrac{1}{2}\kappa T+m^{2}\right)+(p^{\mu}+eA^{\mu})(p^{\nu}+eA^{\nu})-\kappa T^{\mu\nu}\right)\int d^{4}xe^{-ip_{\tau}x^{\tau}}D_{\nu\alpha}(p)
\end{aligned} \qquad (6.5)$$

With terms such as $e^{2}A^{\nu}A^{\mu}$, we immediately recognize that this has the form of the inverse expression for a *non-linear, non-Abelian, Yang-Mills* propagator. Why?

Let us return all the way back to our starting point (1.1), but replace each ordinary derivative $\partial_{;\mu}$ with the $D_{;\mu}$ of (6.3). Thus, (1.1) now becomes:

$$F_{\mu\nu} = D_{;[\mu}A_{\nu]} = D_{;\mu}A_{\nu} - D_{;\nu}A_{\mu} = \partial_{\mu}A_{\nu} - \partial_{\nu}A_{\mu} - ie[A_{\mu},A_{\nu}] = D_{[\mu}A_{\nu]}. \qquad (6.6)$$

For $[A_{\nu},A_{\mu}]=0$, this reduces to (1.1). But, for $[A_{\nu},A_{\mu}]\neq 0$, we recognize (6.6) as being precisely the same as the field strength tensor for a non-Abelian gauge theory. Specifically, if one sets $A_{\nu}\equiv T^{i}A_{i\,\nu}$ and $F_{\mu\nu}\equiv T^{i}F_{i\,\mu\nu}$, where $T^{i}$ are the group generators of $SU(N)$ having a structure relationship $f^{ijk}T_{i} = -i[T^{j},T^{k}]$, with Latin internal symmetry index $i=1,2,3\ldots N^{2}-1$ raised and lowered with the unit matrix $\delta_{ij}$, then (6.6) can be rewritten as the very recognizable non-Abelian field strength tensor:

$$F_{i\,\mu\nu} = \partial_{\mu}A_{i\,\nu} - \partial_{\nu}A_{i\,\mu} - ef_{ijk}A^{j}{}_{\nu}A^{k}{}_{\mu} \qquad (6.7)$$

of Yang-Mills gauge theory. However, the encapsulated relationship $F_{\mu\nu} = D_{[\mu}A_{\nu]}$ of (6.6) is a much preferred, much simpler form in which to represent the field strength. Very importantly, this form the allows the Lagrangian density for a free Yang-Mills field to be expressed, and the integration by parts and the path integral to be calculated, without being split apart into a "perturbative" and "non-perturbative" set of terms which breaks up the gauge symmetry, and without using an artificial "lattice" which ruins the Lorentz symmetry. (See [3], Chapter VII.1 for further discussion of this problem.) Thus, we write:

$$\begin{aligned}
\mathcal{L} &= -\tfrac{1}{4}F^{i}{}_{\mu\nu}F_{i}{}^{\mu\nu} = -\tfrac{1}{2}\mathrm{Tr}(F_{\mu\nu}F^{\mu\nu}) = -\tfrac{1}{2}\mathrm{Tr}(D_{[\mu}A_{\nu]}D^{[\mu}A^{\nu]})\\
&= -\mathrm{Tr}(D_{\mu}A_{\nu}D^{[\mu}A^{\nu]})
\end{aligned}, \qquad (6.8)$$

which precisely mirrors the form of (1.2), but for the trace and the factor of 2, and but for the fact that the ordinary derivatives $\partial_{\mu}$ have been replaced by gauge-covariant derivatives $D_{\mu}$.

Now, the integration-by-parts of (6.8) is trickier than what we did in section 1, because the covariant derivatives do not allow us to apply a "naive" product rule of the form $D_\mu(A_\nu D^{[\mu} A^{\nu]}) = D_\mu A_\nu D^{[\mu} A^{\nu]} + A_\nu D_\mu D^{[\mu} A^{\nu]}$. In fact, it turns out, if one starts with this "naive" product rule as a "hypothesis," then expands out the gauge-covariant derivatives, examines all terms closely, "corrects" this hypothesis accordingly, and re-consolidates everything, that the correct product rule actually is:

$$\partial_\mu(A_\nu D^{[\mu} A^{\nu]}) = D_\mu A_\nu D^{[\mu} A^{\nu]} + A_\nu D_\mu D^{[\mu} A^{\nu]}. \tag{6.9}$$

Here, the derivative in front of the entire term $(A_\nu D^{[\mu} A^{\nu]})$ drops back to an ordinary rather than a gauge-covariant derivative. Therefore, making use of minimal coupling to make this valid in curved spacetime, we rewrite (6.8) as:

$$\begin{aligned}\mathcal{L} &= -\text{Tr}(D_{;\mu} A_\nu D^{;[\mu} A^{\nu]}) = -\text{Tr}(\partial_{;\mu}(A_\nu D^{;[\mu} A^{\nu]})) + \text{Tr}(A_\nu D_{;\mu} D^{;[\mu} A^{\nu]}) \\ &= -\text{Tr}(\partial_{;\mu}(A_\nu D^{;[\mu} A^{\nu]})) + \text{Tr}(A_\mu(g^{\mu\nu} D_{;\sigma} D^{;\sigma} - D^{;\nu} D^{;\mu}) A_\nu)\end{aligned} \tag{6.10}$$

This is the Yang-Mills equivalent of (1.4), and once again, but for the trace and the factor of 2, and but for the fact that the ordinary derivatives $\partial_\mu$ have been replaced by gauge-covariant derivatives $D_\mu$ – and now, but for the fact that the boundary term still contains $\partial_{;\mu}$ and not $D_{;\mu}$, which immensely simplifies the integration of the boundary term using Gauss' / Stokes' theorem – these are identical expressions. The term $g^{\mu\nu} D_{;\sigma} D^{;\sigma} - D^{;\nu} D^{;\mu}$ in the above, if one adds a mass by hand and makes it $g^{\mu\nu}(D_{;\sigma} D^{;\sigma} + m^2) - D^{;\nu} D^{;\mu}$, is *identical* to the term which appears on the top line of (6.4), which caused us to enter this discussion in the first place. This is *why* we stated earlier that (6.4) and (6.5) have the form of the inverse expression for a *non-Abelian, Yang-Mills* propagator.

Now, given that (6.4) has the form of the inverse expression for a non-Abelian, Yang-Mills propagator, what does this actually mean? Have we somehow turned QED into a non-Abelian interaction? Does this mean that the whole path integration of section 2 needs to be re-done and the Green functions recalculated because (2.1) only contained the term $g^{\mu\nu}(\partial_{;\sigma} \partial^{;\sigma} + m^2) - \partial^{;\nu} \partial^{;\mu}$, and not $g^{\mu\nu}(D_{;\sigma} D^{;\sigma} + m^2) - D^{;\nu} D^{;\mu}$? No! And here is why:

Proceed forward to determine an action $S(A^\mu) = \int_U \sqrt{-g}\, d^4x\, \mathcal{L}$ based on (6.10), just as we did in section 1. Because the "trace" in (6.10) is a sum over diagonal elements of a matrix, we can commute the trace with integration at will, thus $\text{Tr}\int = \int \text{Tr}$. So, the action based on (6.10), which is a *Yang-Mills action*, is given by:

$$S(A^\mu) = \int_U \sqrt{-g}\, d^4x\, \mathcal{L}$$
$$\int_U \sqrt{-g}\, d^4x\, \mathcal{L} = -\text{Tr}\int_U \sqrt{-g}\, d^4x\, \partial_{;\mu}\left(A_\nu D^{[\mu} A^{\nu]}\right) + \text{Tr}\int_U \sqrt{-g}\, d^4x\left(A_\mu \left(g^{\mu\nu} D_{;\sigma} D^{;\sigma} - D^{;\nu} D^{;\mu}\right) A_\nu\right). \quad (6.11)$$

We will not worry here about using $R^{\mu\nu} A_\nu = [\partial^{;\nu}, \partial^{;\mu}] A_\nu$ from (1.5) to inject the Ricci tensor and the Einstein equation, though this can also be done, just as in section (1). Once again, using $\partial_{;\mu} V^\mu = (1/\sqrt{-g})\partial_\mu(\sqrt{-g} V^\mu)$ from (1.8), we can turn the $\partial_{;\mu}$ into an ordinary $\partial_\mu$ and then integrate the boundary term,

$$\int_U d^4x\, \partial_\mu\left(\sqrt{-g} A_\nu D^{[\mu} A^{\nu]}\right) = \int_{\partial U} \sqrt{-g}\, (d^3x)_\mu A_\nu D^{[\mu} A^{\nu]}. \quad (6.12)$$

Thus, again giving the gauge field a small mass and adding a source term $J^\nu A_\nu$ with the correct "factor of 2" coefficient just as we did in section (1.10), we arrive at:

$$\boxed{S(A^\mu) = -\text{Tr}\int_{\partial U} \sqrt{-g}\, (d^3x)_\mu A_\nu D^{[\mu} A^{\nu]} + \text{Tr}\int_U \sqrt{-g}\, d^4x\left(A_\mu\left(g^{\mu\nu}\left(D_{;\sigma} D^{;\sigma} + m^2\right) - D^{;\nu} D^{;\mu}\right) A_\nu + 2 J^\nu A_\nu\right)}. \quad (6.13)$$

This is the Yang-Mills counterpart to (1.10), and it must be used in the path integration whenever we have a Yang-Mills field involved. Thus, the Yang-Mills path integral corresponding with (2.1) is:

$$\boxed{\begin{aligned} Z &= \int DA\, e^{iS(A^\mu)} \equiv \mathcal{C}\exp i[W(J)] \equiv \\ &\int DA \exp i\int_U d^4x\, \text{Tr}\left[-\partial_\mu\left(\sqrt{-g} A_\nu D^{[\mu} A^{\nu]}\right) + \sqrt{-g} A_\mu\left(g^{\mu\nu}\left(D_{;\sigma} D^{;\sigma} + m^2\right) - D^{;\nu} D^{;\mu}\right) A_\nu + 2\sqrt{-g} J^\nu A_\nu\right] \end{aligned}}. \quad (6.14)$$

This, of course, will indeed produce terms of third and fourth order in $A_\nu$, see the expanded (6.4), which will cause additional terms to appear in $-V(\phi)$ of (2.6) and therefore introduce third and forth order term using the substitution $\varphi \to -i\delta/\delta(\sqrt{-g} J)$, and therefore make more tedious the derivation of the Green functions as in section 3. But, although more tedious, the calculation is still completely well-defined, and can be done on a deductive basis from (6.14), yielding an exact set of Green functions for Yang Mills theory in curved spacetime. In the process, the "splitting" of perturbative gauge theory is no longer an issue, and we move beyond lattice gauge theory because (6.14) does no violence whatsoever to Lorentz symmetry.

So, need we do this here? Need we recalculate everything in sections 2 and 3 using (6.14)? Or, do the Green functions derived in section 3, and particularly (3.39), still apply here?

It is now important to note that all of the $\partial_{;\sigma} \to D_\sigma$ replacements which appear in (6.13) and (6.14) originate in (6.6), $F_{\mu\nu} = \partial_\mu A_\nu - \partial_\nu A_\mu - ie[A_\mu, A_\nu]$. But, in QED, we have $[A_\mu, A_\nu] = 0$. Therefore, $F_{\mu\nu}$ in (6.6) reduces down to the $F_{\mu\nu}$ in (1.1), and we can proceed precisely as was done in sections 1 through 3 to develop the Green functions of (3.39). This means in particular, that in an Abelian gauge theory, for which $[A_\mu, A_\nu] = 0$, we can replace the gauge-covariant derivatives $D_{;\sigma}$ in all the expressions $g^{\mu\nu}(D_{;\sigma}D^{;\sigma} + m^2) - D^{;\nu}D^{;\mu}$, with ordinary (still differential geometry covariant) derivatives $\partial_{;\sigma}$. That is, for Abelian gauge theory only, because and only because $[A_\mu, A_\nu] = 0$ for QED, we can revert (6.4) back to (6.1), and use (6.1) as is, in order to explicitly calculate the propagator. And, we need make no change whatsoever, for QED, to the Green function and path integral calculation leading to (3.39).

## 7. Explicit Form Derivation of the QED Propagator in the Presence of Matter

As a result of the foregoing, we now use the "ordinary" $\partial^{;\nu}$ in (6.1) to operate on $e^{-ip_\tau x^\tau}$ and so make the substitution $\partial^{;\nu} e^{-ip_\tau x^\tau} \to -ip^\nu e^{-ip_\tau x^\tau}$ throughout, thus arriving at:

$$\mathcal{F}\left(\sqrt{-g(x)}\right)\delta^\mu{}_\alpha = \left(g^{\mu\nu}\left(-p_\sigma p^\sigma + m^2\right) + p^\nu p^\mu\right)\int d^4 x e^{-ip_\tau x^\tau} D_{\nu\alpha}(p)$$
$$= \left(g^{\mu\nu}\left(-p_\sigma p^\sigma + m^2\right) + p^\mu p^\nu + R^{\mu\nu}\right)\int d^4 x e^{-ip_\tau x^\tau} D_{\nu\alpha}(p) \qquad (7.1)$$
$$= \left(g^{\mu\nu}\left(-p_\sigma p^\sigma + \tfrac{1}{2}\kappa T + m^2\right) + p^\mu p^\nu - \kappa T^{\mu\nu}\right)\int d^4 x e^{-ip_\tau x^\tau} D_{\nu\alpha}(p)$$

Then, we restore the delta via $\delta^{(4)}(p) = \mathcal{F}(1) = \int d^4 x e^{-ip_\tau x^\tau}$, to write:

$$\mathcal{F}\left(\sqrt{-g(x)}\right)\delta^\mu{}_\alpha = \left(g^{\mu\nu}\left(-p_\sigma p^\sigma + m^2\right) + p^\nu p^\mu\right)\delta^{(4)}(p) D_{\nu\alpha}(p)$$
$$= \left(g^{\mu\nu}\left(-p_\sigma p^\sigma + m^2\right) + p^\mu p^\nu + R^{\mu\nu}\right)\delta^{(4)}(p) D_{\nu\alpha}(p) \qquad (7.2)$$
$$= \left(g^{\mu\nu}\left(-p_\sigma p^\sigma + \tfrac{1}{2}\kappa T + m^2\right) + p^\mu p^\nu - \kappa T^{\mu\nu}\right)\delta^{(4)}(p) D_{\nu\alpha}(p)$$

We can also use "inverse" notation, to absorb the indexes in $\delta^\mu{}_\alpha$, thus:

$$\delta^{(4)}(p) D_{\nu\alpha}(p) = \mathcal{F}\left(\sqrt{-g(x)}\right)\left(g^{\alpha\nu}\left(-p_\sigma p^\sigma + m^2\right) + p^\nu p^\alpha\right)^{-1}$$
$$= \mathcal{F}\left(\sqrt{-g(x)}\right)\left(g^{\alpha\nu}\left(-p_\sigma p^\sigma + m^2\right) + p^\alpha p^\nu + R^{\alpha\nu}\right)^{-1} \qquad (7.3)$$
$$= \mathcal{F}\left(\sqrt{-g(x)}\right)\left(g^{\alpha\nu}\left(-p_\sigma p^\sigma + m^2\right) + p^\alpha p^\nu + R^{\alpha\nu}\right)^{-1}$$

Now, we may substitute (7.3) into (5.35) to obtain:

$$D_{v\alpha}(x-y) = \int_{U(p)} \frac{1}{(2\pi)^4} d^4p \left( g^{\alpha v}\left(-p_\sigma p^\sigma + m^2\right) + p^v p^\alpha \right)^{-1} e^{ip_\sigma(x-y)^\sigma}$$

$$= \int_{U(p)} \frac{1}{(2\pi)^4} d^4p \left( g^{\alpha v}\left(-p_\sigma p^\sigma + m^2\right) + p^\alpha p^v + R^{\alpha v} \right)^{-1} e^{ip_\sigma(x-y)^\sigma} . \qquad (7.4)$$

$$= \int_{U(p)} \frac{1}{(2\pi)^4} d^4p \left( g^{\alpha v}\left(-p_\sigma p^\sigma + m^2\right) + p^\alpha p^v + R^{\alpha v} \right)^{-1} e^{ip_\sigma(x-y)^\sigma}$$

This, after all is said and done, is just an ordinary inverse Fourier transform. When expressed in terms of $D_{v\alpha}(x-y)$ – which, importantly, is exactly the propagator as it appears in the path integral and Green functions (3.39) – all of the $\sqrt{-g}$ factors and all of the $\delta^{(4)}(p)$ have identically dropped out!

Before proceeding to calculate the inverses in (7.1) through (7.4), there is one other important nuance which we must address. In (6.1), from which (7.1) is immediately derived, and indeed throughout the entire development, we have been dealing with non-commuting derivatives $[\partial^{;v}, \partial^{;\mu}] \neq 0$. This is due to the very nature of differential geometry and parallel transport in curved spacetime, and it starts with the Riemann tensor $R^\alpha{}_{\beta\mu v} A_\alpha = [\partial_{;\mu}, \partial_{;v}] A_\beta$ and works its way onto everything else through $R^{\mu v} A_v = [\partial^{;v}, \partial^{;\mu}] A_v$ of (1.5), and then the Einstein equation written as $R^{\mu v} = -\kappa(T^{\mu v} - \frac{1}{2} g^{\mu v} T)$ just after (1.6). It should therefore not be a surprise, then when we translate into momentum space, non-commuting $\partial^{;v}$ will turn into non-commuting $p^v$, and that is precisely what happens in (7.1) through (7.4).

It is easiest to see this from the first two line of (7.2) which, with everything else factored out, can be reduced down to:

$$\boxed{R^{\mu v} D_{v\alpha}(p) = -\kappa\left(T^{\mu v} - \tfrac{1}{2} g^{\mu v} T\right) D_{v\alpha}(p) = [p^v, p^\mu] D_{v\alpha}(p)}. \qquad (7.5)$$

This important result (which in Yang-Mills gauge theory *may* lead to the consideration of non-commuting $p^v \equiv T^i p_i^v$), precisely mirrors $R^{\mu v} A_v = [\partial^{;v}, \partial^{;\mu}] A_v$ of (1.5), but with the $v$ index on the commutator contracted with the same $v$ index on the propagator $D_{v\alpha}(p)$, and with the other propagator index $\alpha$ having a "free ride." (And, this can be reduced further using spin sums akin to the flat spacetime $\sum_\lambda \varepsilon_v^{(\lambda)} * \varepsilon_\alpha^{(\lambda)} = -g_{v\alpha} + p_v p_\alpha / m^2$, though that is not necessary here.)

Now, with all of the foregoing in mind, we are ready to explicitly calculate the inverse $\left(g^{\alpha\nu}\left(-p_\sigma p^\sigma + m^2\right) + p^\nu p^\alpha\right)^{-1}$. First, to simplify calculation, let us define the an "inverse" $I_{\nu\alpha}$ cleaned of $\delta^{(4)}(p)$ and $\mathcal{F}\left(\sqrt{-g(x)}\right)$, as:

$$I_{\nu\alpha} \equiv \frac{\delta^{(4)}(p)}{\mathcal{F}\left(\sqrt{-g(x)}\right)} D_{\nu\alpha}(p), \tag{7.6}$$

and thereby rewrite (7.2) in simplified form as:

$$\begin{aligned}\delta^\mu{}_\alpha &= \left(g^{\mu\nu}\left(-p_\sigma p^\sigma + m^2\right) + p^\nu p^\mu\right)I_{\nu\alpha} \\ &= \left(g^{\mu\nu}\left(-p_\sigma p^\sigma + m^2\right) + p^\mu p^\nu + R^{\mu\nu}\right)I_{\nu\alpha} \\ &= \left(g^{\mu\nu}\left(-p_\sigma p^\sigma + \tfrac{1}{2}\kappa T + m^2\right) + p^\mu p^\nu - \kappa T^{\mu\nu}\right)I_{\nu\alpha}\end{aligned} \tag{7.7}$$

Note that the same commutation relationship (7.5) applies to $I_{\nu\alpha}$, that is:

$$R^{\mu\nu} I_{\nu\alpha} = -\kappa\left(T^{\mu\nu} - \tfrac{1}{2} g^{\mu\nu} T\right) I_{\nu\alpha} = [p^\nu, p^\mu] I_{\nu\alpha}. \tag{7.8}$$

Now, we already know something about $I_{\nu\alpha}$, because of the usual propagator $D_{\nu\alpha}(p) = \left(-g_{\nu\alpha} + p_\nu p_\alpha / m^2\right)/\left(p_\sigma p^\sigma - m^2\right)$ from $\delta^\mu{}_\alpha = \left(g^{\mu\nu}\left(-p_\sigma p^\sigma + m^2\right) + p^\mu p^\nu\right) D(x-y)_{\nu\alpha}$. Therefore, working with the third line of (7.7), let us use the form:

$$I_{\nu\alpha} = \frac{-g_{\nu\alpha} + \dfrac{p_\nu p_\alpha}{m^2} + X_{\nu\alpha}}{p_\sigma p^\sigma - \tfrac{1}{2}\kappa T - m^2}, \tag{7.9}$$

with $X_{\nu\alpha}$ representing "unknown" additional terms to be deduced. Inserting the above into (7.7) starts us out with:

$$\delta^\mu{}_\alpha = \frac{\left(g^{\mu\nu}\left(-p_\sigma p^\sigma + \tfrac{1}{2}\kappa T + m^2\right) + p^\mu p^\nu - \kappa T^{\mu\nu}\right)\left(-g_{\nu\alpha} + \dfrac{p_\nu p_\alpha}{m^2} + X_{\nu\alpha}\right)}{p_\sigma p^\sigma - \tfrac{1}{2}\kappa T - m^2}. \tag{7.10}$$

The calculation from here is algebraic. After the $\delta^\mu{}_\alpha$ term drops out from either side in the usual way, we remove the denominator, and remove the terms $m^2 g^{\mu\nu} \dfrac{p_\nu p_\alpha}{m^2} - g_{\nu\alpha} p^\mu p^\nu$ which cancel identically in the usual way. We are then left with:

$$\begin{aligned}0 = &\frac{p^\mu p^\sigma p_\sigma p_\alpha - p_\sigma p^\sigma p^\mu p_\alpha}{m^2} - \kappa\left(T^{\mu\nu} - \tfrac{1}{2} T g^{\mu\nu}\right)\frac{p_\nu p_\alpha}{m^2} + \kappa T^{\mu\nu} g_{\nu\alpha} \\ &+ \left(g^{\mu\nu}\left(-p_\sigma p^\sigma + \tfrac{1}{2}\kappa T + m^2\right) + p^\mu p^\nu - \kappa T^{\mu\nu}\right) X_{\nu\alpha}\end{aligned} \tag{7.11}$$

Now, ordinarily, where $[p^\nu, p^\mu] D_{\nu\alpha}(p) = 0$, the first terms $p^\mu p^\sigma p_\sigma p_\alpha - p_\sigma p^\sigma p^\mu p_\alpha$ would also cancel identically, but here, this is not the case. Because of (7.8), we need to commute the $p^\mu$ in $p^\mu p^\sigma p_\sigma p_\alpha$ two terms to the right before we can drop this term. One may note as discussed earlier in connection with (1.5) that $R^{\mu\nu} I_{\nu\alpha} = [p^\nu, p^\mu] I_{\nu\alpha}$ does *not* means that $R^{\mu\nu} = [p^\nu, p^\mu]$ with the $I_{\nu\alpha}$ peeled off, because the index contraction is essential to make this work. But, in (7.11), although not explicit, this contraction is still implicitly in place, because (7.11) it is simply a "downstream" version of (7.7). Thus, after the first commutation using (7.8):

$$0 = \frac{p^\sigma p^\mu p_\sigma p_\alpha - p_\sigma p^\sigma p^\mu p_\alpha}{m^2} + \kappa T^{\mu\nu} g_{\nu\alpha} \\ + \left(g^{\mu\nu}\left(-p_\sigma p^\sigma + \tfrac{1}{2}\kappa T + m^2\right) + p^\mu p^\nu - \kappa T^{\mu\nu}\right) X_{\nu\alpha}$$ (7.12)

dropping out the term with the Einstein equation. After the second commutation, we get:

$$0 = \kappa\left(T^{\mu\sigma} - \tfrac{1}{2}g^{\mu\sigma}T\right)\frac{p_\sigma p_\alpha}{m^2} + \kappa T^{\mu\nu} g_{\nu\alpha} + \left(g^{\mu\nu}\left(-p_\sigma p^\sigma + \tfrac{1}{2}\kappa T + m^2\right) + p^\mu p^\nu - \kappa T^{\mu\nu}\right) X_{\nu\alpha},$$ (7.13)

which eliminates the term with $p^\mu p^\sigma p_\sigma p_\alpha - p_\sigma p^\sigma p^\mu p_\alpha$ while adding back in the Einstein equation term, effective flipping the sign of this term in relation to (7.11).

Now, we simply isolate $X_{\nu\alpha}$ in (7.13), to write, using the "inverse" form:

$$X_{\nu\alpha} = \left(g^{\mu\nu}\left(-p_\sigma p^\sigma + \tfrac{1}{2}\kappa T + m^2\right) + p^\mu p^\nu - \kappa T^{\mu\nu}\right)^{-1}\left(-\kappa\left(T^{\mu\sigma} - \tfrac{1}{2}g^{\mu\sigma}T\right)\frac{p_\sigma p_\alpha}{m^2} + \kappa T^{\mu\nu} g_{\nu\alpha}\right).$$ (7.14)

This is a fascinating result, because we once again have another inverse to determine. But – this is not any inverse: this is *exactly* the same term for which we were seeking the inverse originally, namely, $g^{\mu\nu}\left(-p_\sigma p^\sigma + \tfrac{1}{2}\kappa T + m^2\right) + p^\mu p^\nu - \kappa T^{\mu\nu}$, see our starting point (7.7). This means that the inverse in (7.14) can be replaced by *I*, so long as we simply make sure that the indexes balance. Thus, with the indexes properly balanced, we obtain

$$X_{\nu\alpha} = I_{\nu\mu}\left(-\kappa\left(T^{\mu\tau} - \tfrac{1}{2}g^{\mu\tau}T\right)\frac{p_\tau p_\alpha}{m^2} + \kappa T^{\mu\tau} g_{\tau\alpha}\right).$$ (7.15)

From here, we merely work backwards. Placing (7.15) into (7.9) we obtain:

$$I_{\nu\alpha} = \frac{-g_{\nu\alpha} + \frac{p_\nu p_\alpha}{m^2} + I_{\nu\mu}\left(-\kappa\left(T^{\mu\tau} - \tfrac{1}{2}g^{\mu\tau}T\right)\frac{p_\tau p_\alpha}{m^2} + \kappa T^{\mu\tau} g_{\tau\alpha}\right)}{p_\sigma p^\sigma - \tfrac{1}{2}\kappa T - m^2},$$ (7.16)

giving us a *recursive* expression for $I_{v\alpha}$. Thus, we see that this is not a closed expression, but is rather an infinite series where at each level of recursion, one renames the indexes suitably to avoid confusion, and the substitutes an identical expression back into itself, ad infinitum.

Now, we return to (7.6), and use this in (7.16) to write:

$$\frac{\delta^{(4)}(p)}{\mathcal{F}\left(\sqrt{-g(x)}\right)}D_{v\alpha}(p) = \frac{-g_{v\alpha} + \frac{p_v p_\alpha}{m^2} + \frac{\delta^{(4)}(p)}{\mathcal{F}\left(\sqrt{-g(x)}\right)}D_{v\mu}(p)\left(-\kappa\left(T^{\mu\tau} - \tfrac{1}{2}g^{\mu\tau}T\right)\frac{p_\tau p_\alpha}{m^2} + \kappa T^{\mu\tau}g_{\tau\alpha}\right)}{p_\sigma p^\sigma - \tfrac{1}{2}\kappa T - m^2}.(7.17)$$

Now, as in (7.3), we isolate $\delta^{(4)}(p)D_{v\alpha}(p)$ on the left and substitute this into (5.35), to obtain:

$$D_{v\alpha}(x-y) = \int_{U(p)} \frac{1}{(2\pi)^4}d^4p \frac{-g_{v\alpha} + \frac{p_v p_\alpha}{m^2}}{p_\sigma p^\sigma - \tfrac{1}{2}\kappa T - m^2 + i\varepsilon}e^{ip_\sigma(x-y)^\sigma}$$
$$+ \int_{U(p)} \frac{1}{(2\pi)^4}\frac{\delta^{(4)}(p)}{\mathcal{F}\left(\sqrt{-g(x)}\right)}d^4p \frac{D_{v\mu}(p)\left(-\kappa\left(T^{\mu\tau} - \tfrac{1}{2}g^{\mu\tau}T\right)\frac{p_\tau p_\alpha}{m^2} + \kappa T^{\mu\tau}g_{\tau\alpha}\right)}{p_\sigma p^\sigma - \tfrac{1}{2}\kappa T - m^2 + i\varepsilon}e^{ip_\sigma(x-y)^\sigma}, (7.18)$$

where we have also added the "$+i\varepsilon$" prescription to avert the poles when the integrand reaches $p_\sigma p^\sigma - \tfrac{1}{2}\kappa T - m^2 = 0$. In the term on the first line, we have the usual QED propagator. On the second line, the recursive term is brand new, and it becomes zero when $T^{\mu\tau} = 0$, i.e., in the absence of matter, just as would be expected. In this way, (7.18) describes, from a quantum mechanical standpoint, the effect of matter on the propagation of electromagnetic radiation.

If we denote the gravitational portion of the propagator above as $D_{G\ v\alpha}(x-y)$, such that:

$$D_{G\ v\alpha}(x-y) \equiv \int_{U(p)} \frac{1}{(2\pi)^4}\frac{\delta^{(4)}(p)}{\mathcal{F}\left(\sqrt{-g(x)}\right)}d^4p \frac{D_{v\mu}(p)\left(-\kappa\left(T^{\mu\tau} - \tfrac{1}{2}g^{\mu\tau}T\right)\frac{p_\tau p_\alpha}{m^2} + \kappa T^{\mu\tau}g_{\tau\alpha}\right)}{p_\sigma p^\sigma - \tfrac{1}{2}\kappa T - m^2 + i\varepsilon}e^{ip_\sigma(x-y)^\sigma}, (7.19)$$

then by contrasting to and using (5.34), we can immediately ascertain that:

$$D_{G\ v\alpha}(p) = D_{v\mu}(p)\frac{-\kappa\left(T^{\mu\tau} - \tfrac{1}{2}g^{\mu\tau}T\right)\frac{p_\tau p_\alpha}{m^2} + \kappa T^{\mu\tau}g_{\tau\alpha}}{p_\sigma p^\sigma - \tfrac{1}{2}\kappa T - m^2 + i\varepsilon} = D_{v\mu}(p)\frac{R^{\mu\tau}\left(-g_{\tau\alpha} + \frac{p_\tau p_\alpha}{m^2}\right) + \tfrac{1}{2}\delta^\mu{}_\alpha R}{p_\sigma p^\sigma - \tfrac{1}{2}R - m^2 + i\varepsilon}.(7.20)$$

This is the gravitational interaction portion of the QED momentum space propagator. In this term, the non-linearities, and the interactions between gravitation and electromagnetism, come to the forefront. Because $D_{v\mu}(p)$ is the total propagator in momentum space, the foregoing entails feeding the total $D_{\nu\mu}(p)$ back into the gravitational portion $D_{G\,\nu\alpha}(p)$, which then interacts with terms exclusively involving $T^{\mu\nu}$ and / or $R^{\mu\nu}$ and fully incorporating the Einstein equation $R^{\mu\nu} = -\kappa(T^{\mu\nu} - \tfrac{1}{2}g^{\mu\nu}T)$, in what amounts to an "infinite feedback loop." It is noteworthy that "usual" portion of the propagator in the top line of (7.18) has its simplest appearance, without the factor $\delta^{(4)}(p)/\mathcal{F}(\sqrt{-g(x)})$, when transformed into spacetime, while the gravitational portion in (7.20), from the bottom line of (7.18), discards this factor $\delta^{(4)}(p)/\mathcal{F}(\sqrt{-g(x)})$ when it is transformed into momentum space.

Finally, with the derivation above completes path integration and the Green functions derived in (3.39), because we now have an explicit, albeit infinitely-recursive expression for the propagator. In the path integral itself:

$$Z = \mathcal{C}\exp iW(J(x)) = \mathcal{C}\sum_{n=0}^{\infty} G^{(n)} \frac{1}{2n!}\left(\int_U \sqrt{-g}d^4x\sqrt{-g}d^4y J^{\mu}(x)D_{\mu\nu}(x-y)J^{\nu}(y)\right)^n, \quad (7.21)$$

we simply insert (7.18) with suitable index renaming. For the Green functions,

$$G^{(n)} = \sum_{m=0}^{\infty} (-i)^n \frac{(2n+2m-1)!!}{2m!!}\left(\int_{\partial U}\left(\sqrt{-g}d^3x\sqrt{-g}d^3y\right)_\mu \partial^{:[\mu}D^{\{\nu]}{}_{\nu\}}(x-y)\right)^m, \quad (7.22)$$

for the term $\partial^{:[\mu}D^{\{\nu]}{}_{\nu\}} = 2\partial^{:\mu}D^\nu{}_\nu - \partial_{;\nu}D^{\{\mu\nu\}}$, we need both the symmetric construct $D^{\{\mu\nu\}}$, and the propagator in trace form $D^\nu{}_\nu$. Those too, are readily obtained from (7.18), also using the commutation relationship in (7.5) as appropriate, particularly for $D^{\{\mu\nu\}}$, where the commutation relationship $p^\mu p^\nu + p^\nu p^\mu = -\kappa(T^{\mu\nu} - \tfrac{1}{2}g^{\mu\nu}T) + 2p^\mu p^\nu$ injects the Einstein equation even into the non-gravitational portion (top line of (7.18)) of the propagator.

## 8. Conclusion

We have now shown how to explicitly calculate the QED path integral and associated Green functions, in curved spacetime, with retention of the boundary terms, exactly, and to infinite orders, for any and all spacetime manifolds with sufficient symmetry to admit the application of Pontryagin duality as a form of harmonic analysis. In the process, we have shown, in particular, how gauge symmetry itself greatly facilitates the ability to conduct harmonic analysis in curved spacetime, and to do exact calculations with Pontryagin duality.

There are two directions for further research which become immediately apparent from the foregoing. First, recognizing that Pontryagin duality may overly-restrict the symmetry of the associated spacetime manifold, one should consider repeating the development of sections 4 though 7 using other known techniques for harmonic analysis, or perhaps developing new techniques for harmonic analysis, to see what further results might be achieved. At the same time, the fundamental question of the extent to which path integral quantization can be applied, as is, in curved spacetime and thus allows a union of quantum theory and gravitational theory, or must be viewed as a limiting approximation of something else yet unknown, warrants further discussion and elaboration.

Second, because (6.14) avoids the "Lagrangian splitting" of perturbative gauge theory and avoids lattice gauge theory's compromise of Lorentz / Poincare symmetry, the Yang-Mills path integral (6.14) is extremely important for finding *exact* Green functions and path integrals for the non-Abelian gauge theories of strong and weak interactions. As such, this may provide a necessary foundation for eventually resolving the Yang-Mills "mass gap." [9]